\documentclass[letter,nofootinbib,superscriptaddress,twocolumn]{emulateapj}

\usepackage{amsmath}
\usepackage{amssymb}
\usepackage{braket}
\usepackage{amssymb}
\usepackage{color}
\usepackage{xcolor}
\usepackage{graphicx}
\usepackage{latexsym}
\usepackage{listings}
\usepackage{mathrsfs}
\usepackage{aas_macros}
\usepackage{letltxmacro}
\usepackage[normalem]{ulem}

\usepackage{subfigure}

\usepackage{verbatim}
\usepackage{tabularx}
\usepackage{ragged2e}
\usepackage{multirow}
\usepackage{amsbsy}

\usepackage{numprint}
\usepackage{makecell}
\usepackage[utf8]{inputenc}

\usepackage[colorlinks,citecolor=blue]{hyperref}

\newcommand{\paren}[1]{\left( #1 \right)}
\newcommand{\sqrbrace}[1]{\left[ #1 \right]}

\newcommand{\bhlight}{$\texttt{bhlight}$${}$}
\newcommand{\nubhlight}{$\nu\texttt{bhlight}$${}$}

\newcommand{\fornax}{$\texttt{FORNAX}$${}$}

\newcommand{\detg}{\sqrt{-g}}

\newcommand{\eepsilon}{\epsilon} 
\newcommand{\fin}{f\in \{\nu_e,\nu_{\bar{e}},\nu_x\}}
\newcommand{\sign}{\text{sign}(f)}
\newcommand{\jnuf}{j_{\eepsilon,f}}
\newcommand{\etanuf}{\eta_{\eepsilon,f}}
\newcommand{\Inuf}{I_{\eepsilon,f}}
\newcommand{\chinuf}{\chi_{\eepsilon,f}}
\newcommand{\sigmanuf}{\sigma_{\eepsilon,f}}
\newcommand{\alphanuf}{\alpha_{\eepsilon,f}}
\newcommand{\numin}{\nu_{\text{min}}}
\newcommand{\numax}{\nu_{\text{max}}}

\newcommand{\rhoflr}{\rho_{\text{flr}}}
\newcommand{\uflr}{u_{\text{flr}}}

\newcommand{\lT}{\log_{10}{T}}
\newcommand{\lrho}{\log_{10}{\rho}}
\newcommand{\lrhomin}{\log{\rho}_{\text{min}}}
\newcommand{\lTmin}{\log{T}_{\text{min}}}
\newcommand{\yemin}{(Y_e)_{\text{min}}}
\newcommand{\yemax}{(Y_e)_{\text{max}}}

\newcommand{\Rin}{R_{\text{in}}}
\newcommand{\Rmax}{R_{\text{max}}}

\newcommand{\explain}[1]{}
\newcommand{\added}[1]{#1}

\npdecimalsign{.}
\npthousandsep{,}

\newcolumntype{Y}{>{\RaggedRight\arraybackslash}X}

\sloppypar

\begin{document}

\title{\nubhlight: Radiation GRMHD for Neutrino-Driven Accretion Flows}

\author{Jonah M. Miller}
\email{jonahm@lanl.gov}
\affiliation{Computational Physics and Methods, Los Alamos National Laboratory, Los Alamos, NM, USA}
\affiliation{Center for Theoretical Astrophysics, Los Alamos National Laboratory, Los Alamos, NM, USA}
\affiliation{Center for Nonlinear Studies, Los Alamos National Laboratory, Los Alamos, NM, USA}

\author{Ben. R. Ryan}
\email{brryan@lanl.gov}
\affiliation{Computational Physics and Methods, Los Alamos National Laboratory, Los Alamos, NM, USA}
\affiliation{Center for Theoretical Astrophysics, Los Alamos National Laboratory, Los Alamos, NM, USA}

\author{Joshua C. Dolence}
\email{jdolence@lanl.gov}
\affiliation{Computational Physics and Methods, Los Alamos National Laboratory, Los Alamos, NM, USA}
\affiliation{Center for Theoretical Astrophysics, Los Alamos National Laboratory, Los Alamos, NM, USA}

\begin{abstract}
  The 2017 detection of the in-spiral and merger of two neutron stars
  was a landmark discovery in astrophysics. We now know that such
  mergers are central engines of short gamma ray bursts and sites of
  r-process nucleosynthesis, where the heaviest elements in our
  universe are formed. In the coming years, we expect many more such
  mergers. Modeling such systems presents a significant computational
  challenge along with the observational one. To meet this challenge,
  we present \nubhlight, a scheme for solving general relativistic
  magnetohydrodynamics with energy-dependent neutrino transport in
  full (3+1)-dimensions, facilitated by Monte Carlo methods. We
  present a suite of tests demonstrating the accuracy, efficacy, and
  necessity of our scheme. We demonstrate the potential of our scheme
  by running a sample calculation in a domain of interest---the
  dynamics and composition of the accretion disk formed by a binary
  neutron star merger.
\end{abstract}

\maketitle

\section{Introduction}
\label{sec:intro}



We now know that the in-spiral and merger of two neutron stars is a
central engine of short gamma ray bursts
\cite{GW170817GRB,EichlerBNSGRB,NarayanGRB} and a site of r-process
nucleosynthesis \cite{GW1708MultiMessanger}, where the heaviest
elements in our universe are formed
\cite{LattimerCBC1,LattimerCBC2,CoteRProcess}. In the coming years,
many more such mergers are expected \cite{GW170817PRL}.


This breakthrough poses a number of questions. What are the dynamics
driving the gamma ray burst? Is the relativistic burst of material out
the poles driven by neutrino annihilation \cite{Jaroszynski} or
magnetic fields \cite{BlandfordZnajek}? What fraction of these jets
escapes and what fraction is slowed down by ambient material
\cite{MooleySuperluminal}?  What fraction of the r-process
nucleosynthetic yields comes from material in the tidal tails of the
merging stars and what fraction comes from wind driven off of material
accreting onto the central remnant \cite{TanvirHST}?

This last question is of particular importance for understanding the
spectrum of the optical and infrared afterglow of the merger event
\cite{TanvirHST}. The heavy elements produced via r-process
nucleosynthesis radioactively decay, producing this afterglow---the
so-called \textit{macronova} or \textit{kilonova} \explain{Added
  Metzger (2010) reference}
\added{\cite{Blinnikov,LattimerCBC1,LattimerCBC2,LiPaczynski,MetzgerRProcess,CoteRProcess}}. The
remnant accretion disk ejects mass as a \textit{wind}, which may be
thermal, magnetically-driven, or
neutrino-driven.\footnote{\cite{MetzgerKNReview} and references
  therein offer a nice summary of these processes.} Along with the
tidal tails in the merger, this wind may be one site of r-process
nucleosynthesis.

The dynamics of the r-process in the wind depend on its composition,
which depends on lepton number, and thus neutrino processes and
transport. The mass and morphology of the wind depends on the dynamics
of the disk, which depends on magnetically-driven turbulence via the
magneto-rotational instability \cite{BalbusHawley91}, neutrino
transport, and general relativistic effects such as frame dragging
\cite{wald2010general}. Therefore, accurately computing the
nucleosynthetic yields produced---and thus the spectrum of the
kilonova---depends sensitively on the interplay of gravity, plasma
physics, and neutrino radiation transport. In other words, they are
well-modeled by general relativistic radiation magnetohydrodynamics
(GRRMHD).

Although black hole accretion disk physics is a large and
well-explored topic,\footnote{See \cite{AbramowiczFragileReview} and
  references therein for a review.} very few three-dimensional (3D)
calculations of accretion disks formed by a compact binary merger
including neutrino physics have been performed, and those only
recently. (Indeed, few GRRMHD simulations of disks have been performed
in any context.) \explain{Added Sekiguchi reference}
\added{\cite{SekiguchiBNS} used a hybrid leakage-moment scheme to
  model the radiation in a binary neutron star merger and followed the
  accretion disk formed post-merger.}  \cite{FoucartPostMerger} and
\cite{NouriPostMerger} used a moment method to treat the radiation in
a disk formed by the merger of a black hole and a neutron
star. \cite{SiegelMetzger3DBNS} modeled a disk formed by the merger of
two neutron stars with general relativistic magnetohydrodynamics on a
Cartesian grid and a leakage scheme for the neutrinos.
\cite{SiegelCollapsar} use a similar calculation to argue that
r-process nucleosynthesis can occur in disks formed by the collapse of
massive stars. And \cite{FernandezLongTermGRMHD} performed a suite of
studies of the disk outflow with a cooling function treatment for the
neutrino physics.

These groundbreaking efforts, although heroic, make significant
approximations in the treatment of the radiation transport. Realistic
modeling of neutrino transport requires solving the
$(6+1)$-dimensional kinetic Boltzmann equation for each neutrino
species, which is computationally expensive and numerically
challenging. In the limit of infinite optical depth, the radiation
field may be treated with diffusion physics---see, e.g.,
\cite{GRDiffusion}. For vanishing optical depth, a cooling or leakage
scheme, where neutrinos are allowed to freely stream through material
is appropriate. Indeed this is the approach taken by
\cite{SiegelCollapsar} and \cite{FernandezLongTermGRMHD}. For
intermediate optical depths, the transport equations must be solved
directly. Moment-based schemes, where the continuum limit of the
radiation field is taken and a set of hydrodynamic-like equations are
attained, side-step this requirement by imposing strong assumptions on
the radiation field in order to close the system of equations, with
poorly understood consequences on modeling accuracy.  This approach
was used by \cite{FoucartPostMerger} and
\cite{NouriPostMerger}.\footnote{Much progress has also been made in
  the postprocessing of simulations of accretion disks with realistic
  transport. See, for example, \cite{RichersKasen} and
  \cite{FoucartMCClose}.}


We seek to the remedy this gap. We present \nubhlight, a new GRRMHD
code with accurate neutrino transport via Monte Carlo methods. Monte
Carlo methods solve the full kinetic Boltzmann equation, discretized
with particles, each of which represents a packet of radiation.
\nubhlight\ is built on the successful photon GRRMHD code \bhlight${}$
\cite{bhlight} and is designed specifically to tackle the post-merger
disk problem.

In section \ref{sec:system}, we describe in detail the system of equations
\nubhlight\ is designed to solve. In section \ref{sec:methods}, we
describe the methods used. In section \ref{sec:verification} we
describe code tests used to verify \nubhlight. In section
\ref{sec:disk}, we demonstrate our new capabilities with an example
calculation of a post-merger disk in full 3D with realistic neutrino
transport. Finally, in section \ref{sec:conclusions}, we offer some
concluding thoughts.

\section{System}
\label{sec:system}


We solve the equations of relativistic ideal MHD coupled to neutrino
radiation. We use a formulation almost identical to that presented in
\cite{HARM, grmonty, bhlight}. However, there are a few key
differences. We evolve the conserved lepton number density
(encapsulated in the electron fraction $Y_e$). Unlike \cite{HARM},
this necessitates a realistic, tabulated equation of state. Unlike in
\cite{grmonty,bhlight}, our radiation is relativistic neutrinos, not
photons. Our radiation sector carries conserved lepton number as well
as energy and momentum. Moreover, while there is only one ``type'' of
photon, there are several flavors of neutrinos, which we bundle into
three types: electrons, anti-electrons, and heavies. We discuss the
details of these differences below.

\subsection{Fluid}
\label{sec:system:fluid}

The fluid sector consists of the following system of equations.

\begin{widetext}
\begin{eqnarray}
  \label{eq:particle:cons}
  \partial_t \paren{\detg\rho_0 u^t} + \partial_i\paren{\detg\rho_0u^i}
  &=& 0\\
  \label{eq:energy:cons}
  \partial_t\sqrbrace{\detg \paren{T^t_{\ \nu} + \rho_0u^t \delta^t_\nu}}
  + \partial_i\sqrbrace{\detg\paren{T^i_{\ \nu} + \rho_0 u^i \delta^t_\nu}}
  &=& \detg \paren{T^\kappa_{\ \lambda} \Gamma^\lambda_{\nu\kappa} + G_\nu}\ \forall \nu = 0,1,\ldots,4\\
  \label{eq:mhd:cons}
  \partial_t \paren{\detg B^i} - \partial_j \sqrbrace{\detg\paren{b^ju^i - b^i u^j}} &=& 0\\
  \label{eq:lepton:cons}
  \partial_t\paren{\detg\rho_0 Y_e u^t} + \partial_i\paren{\detg\rho_0Y_eu^i}
  &=& \detg G_{\text{ye}}
\end{eqnarray}
\end{widetext}
where the energy-momentum tensor $T^\mu_{\ \nu}$ is assumed to be
\begin{equation}
  \label{eq:def:Tmunu}
  T^\mu_{\ \nu} = \paren{\rho_0 + u + P + b^2}u^\mu u_\nu + \paren{P + \frac{1}{2} b^2} \delta^\mu_\nu - b^\mu b_\nu
\end{equation}
for metric $g_{\mu\nu}$, rest energy $\rho_0$ fluid four-velocity
$u^\mu$, internal energy density $u$, pressure $P$, and Christoffel connection
$\Gamma^\alpha_{\beta\gamma}$.

Equation \eqref{eq:particle:cons} represents conservation of baryon
number. Equation \eqref{eq:energy:cons} represents conservation of
energy-momentum, subject to the radiation four-force $G_\nu$ (not to
be confused with the Einstein tensor). Note that we have added a
multiple of equation \eqref{eq:particle:cons} to the $\nu = 0$ index
of the canonical form of the energy-momentum equation to arrive at
equation \eqref{eq:energy:cons}. This is equivalent to the canonical
form, but removes rest-energy from the energy conservation law. We
have found this approach to be more numerically favorable. See
\cite{ValenciaGRMHD} for one influential work that uses this trick.

Equation \eqref{eq:mhd:cons}
describes the evolution of magnetic fields, where
\begin{equation}
  \label{eq:def:Bi}
  B^i = ^*F^{it}
\end{equation}
comprise the magnetic field components of the Maxwell tensor
$F_{\mu\nu}$ and $b^\mu$ is the magnetic field four-vector
\begin{equation}
  \label{eq:def:bmu}
  ^*F^{\mu\nu} = b^\mu u^\nu - b^\nu u^\mu.
\end{equation}
Finally, equation \eqref{eq:lepton:cons} describes the conservation of
lepton number. $G_{ye}$ is a source term describing the rate at which
lepton density is transferred between the fluid and the radiation field.
It will be described in more detail below.

The system is closed by an equation of state, which relates the
pressure $P$ to the density $\rho$, internal energy $u$, and and
electron fraction $Y_e$:
\begin{equation}
  \label{eq:def:EOS}
  P = P(\rho, u, Y_e).
\end{equation}
We use an equivalent, temperature-dependent formulation of the
equation of state, which relates the pressure $P$ and specific internal
energy $\varepsilon = u/\rho$ to the density $\rho$, temperature $T$, and
electron fraction $Y_e$:
\begin{eqnarray}
  \label{eq:def:EOS:P:T}
  P &=& P(\rho, T, Y_e)\\
  \label{eq:def:EOS:E:T}
  \frac{u}{\rho} := \varepsilon &=& \varepsilon(\rho, T, Y_e).
\end{eqnarray}
We invert equation \eqref{eq:def:EOS:E:T} to find the temperature and
then calculate the pressure using equation \eqref{eq:def:EOS:P:T}.

\subsection{Neutrino Physics}
\label{sec:system:nu}

\begin{table*}[t!]
  \centering
  \begin{tabular}{l | l | l | l}
    \hline
    \textbf{Type}&\textbf{Processes}&\textbf{Charged/Neutral}&\textbf{Corrections/Approximations}\\
    \hline
    \hline
    Abs./Emis. on Neutrons & 
                             \begin{tabular}{@{}l@{}}
                               $\nu_e + n \leftrightarrow e^- + p$\\
                               $\nu_\mu + n \leftrightarrow \mu^- + p$
                             \end{tabular}
                 & Charged &
                             \begin{tabular}{@{}l@{}}
                               Blocking/Stimulated
                               Abs.\\
                               Weak
                               Magnetism\\
                               Recoil
                             \end{tabular}\\
    \hline
    Abs./Emis. on Protons & 
                            \begin{tabular}{@{}l@{}}
                              $\bar{\nu}_e + p \leftrightarrow e^+ +
                              n$\\
                              $\bar{\nu}_\mu + p \leftrightarrow \mu^+
                              + n$\\
                            \end{tabular}
                 & Charged &
                             \begin{tabular}{@{}l@{}}
                               Blocking/Stimulated
                               Abs.\\
                               Weak
                               Magnetism\\
                               Recoil
                             \end{tabular}\\
    \hline
    Abs./Emis. on Ions & $\nu_eA \leftrightarrow A'e^-$ & Charged &
                                                                    \begin{tabular}{@{}l@{}}
                                                                      Blocking/Stimulated Abs.\\
                                                                      Recoil
                                                                    \end{tabular}\\
    \hline
    Electron Capture on Ions & $e^- + A \leftrightarrow A' + \nu_e$ &
                                                                      Charged &
                                                                                \begin{tabular}{@{}l@{}}
                                                                                  Blocking/Stimulated Abs.\\
                                                                                  Recoil
                                                                                \end{tabular}\\
                                                             
    \hline
    $e^+-e^-$ Annihilation & $e^+e^- \leftrightarrow \nu_i\bar{\nu}_i$&
                                                                    Charged
                                                                    + Neutral &
                                                                    \begin{tabular}{@{}l@{}}
                                                                      single-$\nu$
                                                                      Blocking\\
                                                                      Recoil
                                                                    \end{tabular}\\
    \hline
    $n_i$-$n_i$ Brehmsstrahlung & $n_i^1 + n_i^2 \to n_i^3 +
                                      n_i^4 + \nu_i\bar{\nu}_i$ 
                                    & Neutral & 
                                                \begin{tabular}{@{}l@{}}
                                                  single-$\nu$ Blocking\\
                                                  Recoil
                                                \end{tabular}\\
    \hline
    
  \end{tabular}
  \caption{Emission and Absorption Processes used in \nubhlight.}
  \tablecomments{The
    symbols in the processes are as follows: $n$ is a neutron, $p$ a proton, $e^-$ an
    electron, $e^+$ a proton, $\mu^-$ a muon,
    $\mu^+$ an antimuon, and $n_i$ an arbitrary nucleon. $\nu_i$ is an
    arbitrary neutrino. $\nu_e$ is an electron neutrino, and
    $\bar{\nu}_e$ is an electron antineutrino.
    We describe the corrections and approximations used below, as
    tabulated in \cite{fornax} and provided to us in
    \cite{BurrowsCorresp}.  Blocking and stimulated absorption are
    related to the Fermi-Dirac nature of neutrinos. Weak magnetism is
    related to the extended quark structure of nucleons. Recoil is the
    kinematic recoil. Single-$\nu$ blocking is a Kirkhoff's law based
    approximation of blocking that becomes exact for processes that
    involve only a single neutrino. The details of these interactions
    are summarized in \cite{BurrowsNeutrinos}.}
  \label{tab:emis:abs}
\end{table*}

We are interested in r-process nucleosynthesis, which depends on the
fraction of free neutrons in our gas, or
$$1-Y_e$$
The electron fraction $Y_e$
is affected by the emission or absorption of electron neutrinos
(denoted $\nu_e$) and electron antineutrinos (denoted $\bar{\nu}_e$)
but unaffected by the emission and absorption of all other
neutrinos. We therefore dub the other neutrinos, which do not modify
electron number \textit{heavy} neutrinos and denote them $\nu_x$. If
neutrino species does not matter (or we wish to iterate over species,
depending on context), we denote the neutrinos as $\nu_i$.

We include many different interactions of neutrinos with matter. We
categorize these processes as ``absorption or emission'' and as
``scattering'' processes. We list the absorption and emission
processes in table \ref{tab:emis:abs}. Those that involve the absorption or
emission of an electron neutrino or antineutrino can change the electron
fraction and therefore the number of free neutrons in the gas. We
include the elastic scattering processes listed below,
\begin{eqnarray}
  \label{eq:nui:p}
  \nu_i + p &\leftrightarrow& \nu_i + p\\
  \label{eq:nui:n}
  \nu_i + n &\leftrightarrow& \nu_i + n\\
  \label{eq:nui:a}
  \nu_i + A &\leftrightarrow& \nu_i + A\\
  \label{eq:nui:alpha}
  \nu_i + \alpha &\leftrightarrow& \nu_i + \alpha
\end{eqnarray}
where $n$ represent
neutrons, $p$ protons, $\nu_i$ neutrinos of arbitrary type,
$A$ heavy ions, and $\alpha$ alpha particles.

There are several effects which we are neglecting, mainly inelastic
scattering of neutrinos off of electrons \cite{Bruenn1985nu};
neutrino-neutrino annihilation, which may help drive the gamma ray
burst \cite{eichler1989nucleosynthesis}; and neutrino oscillations
\cite{DuanSNnuOscillations}. We also neglect ion screening, electron
polarization, and form factor corrections to neutrino-heavy ion
scattering \eqref{eq:nui:a}. Although this effect is subdominant in
core-collapse supernovae \cite{BruennMezzacappa1997ion}, we do not
know how important it is for the disk problem. We also note that the
pair processes, i.e., nucleon-nucleon bremsstrahlung and
particle-antiparticle annihilation, do not impose any conditions on
pairs of Monte Carlo radiation packets. Rather, they are approximated
as isotropic processes and incorporated into our emissivities and
absorption opacities. On large scales, i.e., $G M_{BH}/c^2$ for a
black hole of mass $M_{BH}$, we believe this is a good approximation.
Since this work is a ``first pass'' at accurately tracking neutrino
physics in neutrino driven accretion flows, we believe ignoring these
effects initially is justified. We will incorporate and study them in
future work.

Neutrino interactions with matter have a long history in astrophysics
\cite{Freedman1974nu,TubbsSchramm1975nu,FullerFowlerNewman1982nu,Bruenn1985nu,LEINSON198880,Aufderheide1994nu,Horowitz1997nu}. We
borrow these results to produce our emissivities, opacities, and cross
sections. Our emissivities and opacities in particular are drawn from
tabulated data first presented in \cite{BurrowsNeutrinos}, which also
accounts for subdominant high-density many-body effects. Scattering is
treated on an interaction-by-interaction basis and we use analytic
single-particle cross-sections. We use cross-sections as summarized in
\cite{BurrowsNeutrinos}.

\subsection{Treatment of the Neutrino Radiation Field}
\label{sec:system:rad}

We assume our neutrinos are massless and travel on null geodesics and
obey a light-like dispersion relation
\begin{equation}
  \label{eq:nu:dispersion:relation}
  -k^\mu \eta_\mu = \eepsilon = h\nu,
\end{equation}
where $h$ is Planck's constant, $\eepsilon$ is the energy of a
neutrino with wavevector $k^\mu$ as measured by an observer traveling
along a timelike Killing vector $\eta^\mu$. Here $\nu$ is the
frequency of the neutrino. However, to avoid notational confusion, we
will usually use $\eepsilon$ rather than $\nu$ when referring to
neutrino energies and frequencies, which are interchangeable via a
factor of Planck's constant. Since the neutrino mass is both small
and unknown---far smaller than the many-MeV energies neutrinos attain
in post-merger disks---we believe this is a reasonable approximation.

We thus recast our neutrino transport as the standard radiative
transfer equation
\begin{equation}
  \label{eq:radiative:transfer}
  \frac{D}{d\lambda}\paren{\frac{h^3\Inuf}{\eepsilon^3}} = \paren{\frac{h^2\etanuf}{\eepsilon^2}} - \paren{\frac{\eepsilon \chinuf}{h}} \paren{\frac{h^3\Inuf}{\eepsilon^3}},
\end{equation}
where $D/d\lambda$ is a derivative along a neutrino trajectory in
phase space, $\Inuf$ is the intensity of the neutrino field of
flavor $f\in \{\nu_e,\bar{\nu}_e,\nu_x\}$,
\begin{equation}
  \label{eq:def:extinction:coeff}
  \chinuf = \alphanuf + \sigmanuf^a
\end{equation}
is the extinction coefficient that combines absorption coefficient
$\alphanuf$ and scattering extinction $\sigmanuf^a$ for scattering
interaction $a$ and
\begin{equation}
  \label{eq:def:emission:coeff}
  \etanuf = \jnuf + \etanuf^s(\Inuf)
\end{equation}
is the emissivity combining fluid emissivity $\jnuf$ and emission due
to scattering from $\etanuf^s$. Note that every neutrino flavor has
its own radiation field and interactions with matter. (Equivalently,
the radiation field has an extra, discrete index specifying neutrino
flavor.) Each of the quantities in expression
\eqref{eq:radiative:transfer} is invariant.


\subsection{Radiation-Fluid Interactions}
\label{sec:system:rad:fluid:coupling}

We define an orthonormal tetrad\footnote{Here roman Greek indices
  indicate a lab frame and latin indices in parentheses indicate a
  comoving frame.}
$$
e^\mu_{(a)}
$$
with
\begin{equation}
  \label{eq:orth:of:tetrad}
  e^\mu_{(a)} e_\nu^{(b)} = \eta^\mu_\nu \eta^{(a)}_{(b)}
\end{equation}
so that
\begin{equation}
  \label{eq:def:tetrad}
 e^\mu_{(t)} = u^\mu,
\end{equation}
i.e., so that it is comoving with the fluid. In this frame, the radiation four-force is
\begin{equation}
  \label{eq:def:G:comoving}
  G_{(a)} = \frac{1}{h}\int d\eepsilon d\Omega \paren{\chinuf \Inuf - \etanuf}n_{(a)},
\end{equation}
where $n_{(a)} = p_{(a)}/\eepsilon$. A coordinate transformation
then maps the comoving radiation four-force into the lab frame:
\begin{equation}
  \label{eq:coordinate:transform}
  G^\mu = e^\mu_{(a)} G^{(a)}.
\end{equation}

The scalar source term $G_{ye}$ for lepton conservation
\eqref{eq:lepton:cons} is similar. Evaluated in the fluid
frame,
it is given by
\begin{equation}
  \label{eq:def:Gye}
  G_{ye} = \frac{m_p}{h}\sign \int \frac{\chinuf \Inuf - \etanuf}{\eepsilon}d\Omega d\eepsilon,
\end{equation}
where $m_p$ is the mass of a proton and
\begin{equation}
  \label{eq:def:neutrino:selector:function}
  \sign = \begin{cases}
    1&\text{if }f = \nu_e\\
    -1&\text{if }f = \bar{\nu}_e\\
    0&\text{if }f = \nu_x
  \end{cases}
\end{equation}
determines the sign of the contribution.

\section{Methods}
\label{sec:methods}

\subsection{Fluid Integration}
\label{sec:meth:fluid}

We evolve our fluid via a standard second-order conservative
high-resolution shock capturing finite volume method. We base our implementation in this
sector on HARM \cite{HARM} and use the same set of primitive and
conserved variables as described in \cite{HARM} and
\cite{bhlight}. The lone exception being the electron fraction. We
describe our implementation of the electron fraction in more detail in
section \ref{sec:passive:scalars}.

We use a local Lax-Friedrichs (LLF) approximate Riemann solver
\cite{HLL}. For reconstructions, we use either a linear reconstruction
with a monotized central slope limiter \cite{toro2013riemann} or a
fifth-order WENO reconstruction \cite{WENO}. The form we use is the
variant described in \cite{sashaWENO5}, although we do not manually
reduce the order of reconstruction near discontinuities.

We treat our magnetic fields via a constrained transport method
described by Toth \cite{TothCT}. This version of constrained transport
uses cell-centered magnetic fields. We use a special second-order
derivative operator which ensures that a discrete, corner-centered divergence of the magnetic
field vanishes. Since this scheme uses centered-differencing, it
neglects emf upwinding, which can be important for flux loop
advection. For more details, see \cite{HARM}.

In general relativity, the conversion between conserved variables and
primitive variables is not known analytically and involves the
numerical root finding of a complex algebraic function. We use the
procedure described by Mignone and McKinney in
\cite{MignoneMcKinneyU2PRIM}.


%


\subsection{Radiation Transport}
\label{sec:meth:rad}


We have modified the way \bhlight$\ $ performs radiation transport. We
transport three types of radiation packet, each one corresponding to
electron neutrinos, electron antineutrinos, or heavy neutrinos. Each
type of neutrino has separate emissivities and opacities. This implies
that the probability that a given radiation packet is emitted,
scattered, or absorbed depends on the neutrino type. There are two
ways we could account for this:
\begin{enumerate}
\item \label{en:separate} Treat each neutrino type as a separate class
  of radiation object and draw probabilities from completely separate
  probability distributions, one for each neutrino type.
\item \label{en:joint} Draw probabilities from a joint probability
  distribution, which depends on neutrino type.
\end{enumerate}
The former allows for more fine-grained control over how phase space
is sampled, while the latter has the advantage of being simpler to
implement. Because of its simplicity, we have implemented option
\ref{en:joint}. We now describe this approach in more detail. Our
treatment closely follows that in \cite{grmonty} with a few
modifications. Here we emphasize the differences between our algorithm
and that described in \cite{grmonty} and \cite{bhlight}.

\subsubsection{Emissivity}
\label{sec:meth:rad:jnu}

The probability distribution of emitted radiation packets is given by
\begin{equation}
  \label{eq:rad:sph:prob}
  \frac{1}{\sqrt{-g}}\frac{dN_s}{d^3x dt d\nu d\Omega}
  = \frac{1}{w\sqrt{-g}}\frac{dN}{d^3x dt d\nu d\Omega}
  = \frac{1}{\omega}\frac{\jnuf}{h\nu},
\end{equation}
where $N_s$ is the number of ``superneutrinos,'' or radiation packets
with $w$ physical neutrinos per packet, $N$ is the number of physical
neutrinos, $\jnuf$ is the emissivity (in the plasma frame) of
neutrinos of species $i$ with de Broglie frequency $\nu$ and flavor
$f\in \{\nu_e,\bar{\nu}_e,\nu_x\}$, and $h$ is Planck's constant.

This implies that the number of superneutrinos created in a time
interval $\Delta t$ is
\begin{equation}
  \label{eq:num:sph:created}
  N_{s,tot} = \Delta t\sum_{\fin}\int \sqrt{-g}d^3xd\nu d\Omega\frac{1}{w}\frac{\jnuf}{h\nu},
\end{equation}
and the number of superneutrinos of flavor $f$ created in a finite
volume cell $i$ of volume $\Delta^3 x$ is given by
\begin{equation}
  \label{eq:numn:sph:created:zone}
  N_{s,i,f} = \Delta t\Delta^3 x \int \sqrt{-g}d\nu d\Omega\frac{1}{w}\frac{\jnuf}{h\nu}.
\end{equation}
We approximately fix the total number of superneutrinos created per
timestep by setting the weights $w$.

We set the weight to
\begin{equation}
  \label{eq:def:sph:weight}
  w = \frac{C}{\nu}
\end{equation}
where $C$ is a constant.

This ensures
that a superneutrino of frequency $\nu$ and $w(\nu)$ has a
total energy
\begin{equation}
  \label{eq:E:superneutrino}
  E_s = w h\nu = h C;
\end{equation}
superneutrino energy is independent
of frequency.

We calculate the constant $C$ by fixing the total number of
superneutrinos created to be $N_{target}$ and inverting equation
\eqref{eq:num:sph:created} for $C$. We decide $N_{target}$ by
  trying to keep the total number of superneutrinos constant over
  time. This means $N_{target}$ is chosen so that the superneutrinos
  created and scattered replace those lost to absorption or that leave
  the domain. This results in the integral quantity
\begin{equation}
  \label{eq:calculate:wgt:C:true}
  C = \frac{\Delta t}{h N_{target}}\sum_{\fin}\int \sqrt{-g}d^3xd\nu d\Omega \jnuf .
\end{equation}

To summarize, to produce superneutrinos, we sample them from a
species dependent probability distribution which has a weight
calculated by integrating over the total probability distribution for
all species.

\subsubsection{Absorption}
\label{sec:meth:rad:abs}

Our treatment of absorption is identical to that described in
\cite{bhlight}, except that absorption extinction coefficients are now
evaluated per neutrino species. Absorption is treated
probabilistically. If a radiation packet of neutrino flavor $f$
travels an affine distance $\Delta \lambda$, it passes through an
incremental optical depth to absorption
\begin{equation}
  \label{eq:rad:dtau:abs}
  \Delta\tau_a(\nu,f) = \nu\alphanuf\Delta\lambda,
\end{equation}
where $\alphanuf$ is the absorption extinction coefficient for
neutrino radiation of flavor $f$ and frequency $\nu$. An absorption
event occurs if
\begin{equation}
  \label{eq:rad:abs:event}
  \Delta\tau_a(\nu,f) > -\ln(r_a),
\end{equation}
where $r_a$ is a random variable sampled uniformly from the interval
$[0,1)$.

\subsubsection{Scattering}
\label{sec:meth:rad:scatt}

Like in \bhlight, scattering in \nubhlight${}$ is treated
probabilistically. We generalize the approach in \bhlight${}$ to treat
scattering of radiation off of multiple scattering particles, each
with their own cross-section. We allow our neutrinos to scatter
elastically off of protons \eqref{eq:nui:p}, neutrons
\eqref{eq:nui:n}, heavy nuclei \eqref{eq:nui:a}, and alpha particles
\eqref{eq:nui:alpha}. We calculate the individual number densities of
the constituent particles via the appropriate mass fraction, which is
tabulated in our equations of state.

For each neutrino flavor $f$ and type
$p$ of gas particle off of which a neutrino can scatter, we
construct a scattering extinction coefficient
$\alpha_s(\nu,f,p)$. Then a superneutrino of flavor $f$ scatters off
of a particle of species $p$ if
\begin{equation}
  \label{eq:rad:scatt:event}
  \Delta \tau_p(f) > -\ln(r_s)/b_s(\nu,f,p),
\end{equation}
where $\Delta \tau_p(f)$ is the scattering optical depth due to an
interaction between the neutrino and the particle, constructed
analogously to the absorption optical depth \eqref{eq:rad:dtau:abs},
and $b_s(\nu,f,p)$ is a bias parameter that enhances the probability
of scattering. To ensure that the biased process reflects nature, we
reset the weight of the scattered superneutrino to $w/b$ for a
conservative process with incident superneutrino of weight $w$. For
more details, see \cite{grmonty} and \cite{bhlight}.

How do we sample multiple different interactions? After all, the
neutrino should be subject to absorption and scattering against all
kinds of particles. When a superneutrino travels an affine distance of
$\Delta \lambda$, we construct all optical depths
\begin{displaymath}
  \{\Delta\tau_i\}_{i\in\{a,p\}}
\end{displaymath}
and biases
\begin{displaymath}
  \{b_i\}_{i\in\{a,p\}}
\end{displaymath}
for absorption and scattering against all particles. Then we sample a
uniform random variable
\begin{displaymath}
  r_i,\ i\in\{a,p\}
\end{displaymath}
for each type of interaction. The interaction that occurs is the one
for which the ratio
\begin{equation}
  \label{eq:interaction:ratio}
  -\frac{\ln(r_i)}{b_i\Delta\tau_i}
\end{equation}
is smallest for all $i\in\{a,p\}$.

Unlike in \cite{grmonty} and \cite{bhlight}, we allow our bias
parameters to depend individually on both the neutrino flavor $f$ and
the species of interacting particle $p$. In particular, we demand that
the parameter
\begin{equation}
  \label{eq:scat:bias:cond}
  b_s(\nu,f,p) \Delta \tau_p(f)
\end{equation}
be approximately equal for all scattering processes. This ensures that
all scattering processes are equally well sampled. In section
\ref{sec:ver:scatt}, we provide evidence that this procedure is both
necessary and effective.

\subsubsection{Sampling The Scattered Superneutrino}
\label{sec:meth:scatt:sample}

To generate a new superneutrino from an incident one with wavevector
$k^\mu$, we follow a modified version of the procedure presented in
\cite{grmonty, bhlight}:
\begin{enumerate}
\item We boost into the rest frame of the plasma.
\item We sample the four-momentum $p^\mu$ of the particle off of which
  the superneutrino scatters from a thermal relativistic Maxwell
  distribution using the procedure described in \cite{Canfield}. Note
  that this requires the differential single-particle cross section
  for the interaction of the scattering particle with the neutrino.
\item We boost into the rest frame of the scattering particle.
\item We sample the wavevector $k^\mu_s$ of the scattered
  superneutrino from the differential single-particle cross section.
\item We transform $k^\mu_s$ back into the lab frame.
\end{enumerate}
This is why we must perform scattering on a per scatterer
basis. Otherwise, the differential cross section is inaccessible.

\subsubsection{Radiation Force}
\label{sec:meth:rad:force}

We calculate the radiation four-force on the fluid by conserving
four-momentum, in a manner identical to that described in
\cite{bhlight}. The only complication emerges from the tracking of
lepton number.

\subsubsection{Tracking Lepton Number}
\label{sec:meth:rad:leptons}

When a superneutrino is emitted or absorbed,
it can modify the electron fraction of the
gas. We couple this contribution to the electron fraction evolution via an operator split update analogous to the radiation four-force update in \cite{bhlight}. The emission of a neutrino radiation
packet of flavor $f$ and weight $w$ provides a discrete contribution
to source term \eqref{eq:def:Gye} of magnitude
\begin{equation}
  \label{eq:ye:source:term:emiss}
  \frac{\Delta (\sqrt{-g}\rho u^0 Y_e)}{\Delta t} = - \sqrt{-g}\frac{w u^0 m_b}{\sqrt{-g} \Delta^4 x}\sign,
\end{equation}
where $u^0$ is the time component of the fluid four-velocity, $m_b$ is
unit mass per baryon in the gas, and $\sqrt{-g}\Delta^4 x$ is the
invariant four-volume of a discrete finite volumes cell (including
time step).  The contribution to this source term for absorption is
equal and opposite.

\subsubsection{A Note on Timesteps}
\label{sec:meth:rad:timestep}

We pause briefly to note limits on the size of our timestep. Our
method is fully explicit, both in the radiation and fluid sectors. We
feel comfortable applying a fully explicit approach because the
systems we are interested in have modest optical depths, and cooling
times are long. Moreover, our approach to scattering and absorption
requires that a superneutrino not travel more than one cell distance in one
timestep. This timestep restriction means that the restriction on
timestep due to using a fully explicit approach is not so severe.

We therefore insist our timestep is smaller than the following
quantities:
\begin{enumerate}
\item The light crossing time within any cell
\item The cooling time due to emissivity $u/\int d\nu d\Omega j_\nu$
\end{enumerate}
The emissivity condition (2) isn't a guarantee for stability. Rather,
it is a guarantee that the system will converge to a stable solution
in the limit of large superneutrino number. In practice we find that
the light crossing time within a cell is almost always the smallest of
these quantities and the cooling time restriction is not severe.

\subsection{Advection and Passive Scalars}
\label{sec:passive:scalars}

As discussed in section \ref{sec:system}, the electron fraction $Y_e$
evolves via equation \eqref{eq:lepton:cons}. We have implemented a
generic framework for evolving variables which are ``passively''
advected by the fluid, so called \textit{passive variables.} Our
framework allows for two methods of advection, which differ by what
variable is considered the primitive variable. We dub these two
approaches \textit{advect intrinsics} and \textit{advect numbers}.

\subsubsection{Advecting Intrinsics}
\label{sec:passive:intrinsics}

An \textit{extrinsic} thermodynamic quantity is one which grows
linearly with volume. Extrinsic quantities include total energy and
entropy. An \textit{intrinsic} quantity is the corresponding extrinsic
quantity \textit{per volume}. This is in contrast to, say,
\textit{specific} quantities, which are extrinsic quantities
\textit{per mass}.

Consider an intrinsic variable $\phi$ that is passively advected by
the fluid. It obeys the differential equation
\begin{equation}
  \label{eq:passive:intrinsic}
  (\phi u^\mu)_{;\mu} = 0,
\end{equation}
where $u^\mu$ is the fluid four-velocity. If we perform a $(3+1)$
split and translate this into the language of finite volumes, $\phi$
is our primitive variable, $\phi u^0$ is our conserved variable, and
$\phi u^i$ is our flux in the $i^{th}$ direction. If
\begin{equation}
  \label{eq:Ye:to:intrinsic}
  \phi = \rho Y_e,
\end{equation}
then we recover equation \eqref{eq:lepton:cons} for conservation of
lepton number.

\subsubsection{Advecting Numbers}
\label{sec:passive:numbers}

Consider a ``number'' quantity $X$, which is neither intrinsic nor
extrinsic. The ``density'' $X\rho$ will be intrinsic. This fact
suggests an alternative form of equation \eqref{eq:passive:intrinsic}:
\begin{equation}
  \label{eq:passive:number}
  (X\rho u^\mu)_{;\mu} = 0,
\end{equation}
where the conserved variable is now $X\rho u^0$ and the $i^{th}$ flux
is $X\rho u^i$. However, there is a degeneracy in the primitive
variable. If we treat $X\rho$ as the primitive, we recover the
formalism in section \ref{sec:passive:intrinsics}. If we treat $X$ as
the primitive, we recover a mathematically equivalent but numerically
distinct approach. This is the \textit{advect numbers} scheme.

If $X = Y_e$, then equation \eqref{eq:passive:number} becomes equation
\eqref{eq:lepton:cons} for the conservation of lepton number. In
practice, we have found that advecting $Y_e$ as a number as per
equation \eqref{eq:passive:number} to be more numerically
robust---particularly in the atmosphere---than advecting the conserved
proton mass density as per equation \eqref{eq:passive:intrinsic} and
this is the approach we use.

\subsection{Equation of State}
\label{sec:meth:eos}

We have implemented realistic, tabulated nuclear equations of
state. We use the tables as generated and described in
\cite{stellarcollapsetables,stellarcollapseweb}. Our equation of
state tables provide thermodynamic quantities in terms of the log of the
density, $\lrho$, the log of the temperature $\lT$, and
the election fraction $Y_e$.

Since temperature is not one of our primitive (or conserved)
variables, we must solve for it via one-dimensional root finding. In
our implementation, we use Newton's method but default to bisection if
Newton's method fails. To do so, we invert the relation
\begin{equation}
  \label{eq:def:specific:internal:energy}
  \rho \varepsilon(\lrho,\lT,Y_e) = u
\end{equation}
with a given specific internal energy $\varepsilon$ or the relation
\begin{eqnarray}
  \label{eq:def:enthalpy}
  \rho &=& w(\lrho,\lT,Y_e) \\
       &&\quad - \rho \varepsilon(\lrho,\lT,Y_e)  \nonumber\\
       &&\qquad - P(\lrho,\lT,Y_e) \nonumber
\end{eqnarray}
with a given enthalpy by volume $w$ for $\lT$, where $\lrho$ and
$Y_e$ are given by the primitive state. We can then extract
thermodynamic quantities such as pressure and sound speed. (We use
equation \eqref{eq:def:enthalpy} when solving for our primitive
variables from our conserved variables and
\eqref{eq:def:specific:internal:energy} everywhere else.)

\subsection{Atmosphere Treatment}
\label{sec:meth:atmo}

Equations \eqref{eq:particle:cons}, \eqref{eq:energy:cons}, and
\eqref{eq:lepton:cons} are valid only for non-vanishing density
$\rho$. Moreover, only some values of internal energy $u$ and electron fraction
$Y_e$ are physically valid. Therefore, we must impose floors on these
quantities to keep them in a physically valid regime.

An additional complication is that tabulated thermodynamic
values are available only for a finite range of temperature, pressure,
and electron fraction:
\begin{eqnarray}
  \label{eq:tab:range}
  \log_{10}\rho &\in& [\lrhomin,\text{log}\rho_{\text{max}}]\nonumber\\
  \log_{10}T &\in& [\lTmin,\text{log}T_{\text{max}}]\\
  \text{and }Y_e &\in& [\yemin,\yemax].\nonumber
\end{eqnarray}
Therefore these limits must be accounted for in some way. The
physically allowed values of electron fraction sit well within the
range given by equation \eqref{eq:tab:range}, so we simply set floors
and ceilings for the electron fraction given by $\yemin$ and $\yemax$.

For density in black hole metrics, we demand that
\begin{equation}
  \label{eq:density:floor}
  \rho >= \rhoflr = \frac{\rho_0}{r^2}
\end{equation}
where we choose $\rho_0 \approx 10^{-5}$
for our disk simulations.\footnote{This
  choice is problem dependent.}
This implies that near the black hole,
our floor is about $10^{-5}$ in code units.\footnote{Typically,
  approximately $10^5$ g$/$cm$^3$ in physical units.} However, the
floor decays in radius, so it is much smaller at large radii. This
treatment is designed to ensure that the atmosphere does not interfere
with diffuse winds blown off the disk. 

For these large radii, $\rhoflr < \lrhomin$. We therefore analytically
extend the table with a cold, polytropic equation of state (which
depends on electron fraction),
\begin{equation}
  \label{eq:def:polytrope}
  P_{\text{poly}} = K(Y_e)\rho^{\Gamma(Y_e)},
\end{equation}
where $K$ and $\Gamma$ are chosen so that $P_{\text{poly}}$ and
$(\partial P_{\text{poly}}/\partial\rho)_s$ match the table at
$(\lrhomin,\lTmin,Y_e)$.

For internal energy $u$, we demand that
\begin{equation}
  \label{eq:u:floor}
  u >= \uflr(Y_e) = \rhoflr\varepsilon(\rhoflr,\lTmin,Y_e),
\end{equation}
where the specific internal energy $\varepsilon$ can contain
contributions from binding energy and thus be negative. For
consistency with the cold nature of the polytropic equation of state
\eqref{eq:def:polytrope}, we set $u = \uflr$ when $\rho$ is less than
some threshold value, typically a $\lrhomin/10^5$ or less. In
magnetically dominated regions, these floors are imposed in the rest
frame of the fluid. However, in matter or kinetic energy dominated
regions, they are imposed in the lab frame.

We note that our treatment of the floors is not thermodynamically
consistent. Moreover, any application of density floors is unphysical
and can, if care is not taken, change the results of a
simulation. Unfortunately, within an Eulerian framework, we have no
choice but to apply density floors. Fortunately, this inconsistency
affects only very low density regions and should therefore not change
the results of a simulation if used judiciously. We have experimented with floor values
and found the results of our simulations to be insensitive to these
choices.

Electron fraction has no meaning in the atmosphere, but numerically,
we must set it to something. For simplicity, we set the atmosphere to
have an electron fraction of $Y_e=0.5$ at the initial time. When
density floors are enforced, $Y_e$ is not reset. Rather, the bounds on
$Y_e$ are enforced independently.

\subsection{Tracer Particles}
\label{sec:meth:tracers}

We have added tracer particles to \nubhlight. A tracer particle is a
numerical representation of a Lagrangian fluid packet, which is
passively advected with the fluid. In the (3+1)-split of general
relativity, tracers obey the equation of motion
\begin{equation}
  \label{eq:tracer:eom}
  \frac{d x^i}{dt} = \frac{u^i}{u^0} = \alpha v^i - \beta^i,
\end{equation}
where $x^i$ are the spatial components of the tracer's position
vector, $\alpha$ is the lapse, $\beta^i$ the components of the shift,
$v^i$ the three-velocity of the fluid, and $u^\mu$ the four-velocity
of the fluid \cite{Foucart2014BHNS}.

We interpolate the velocities in equation \eqref{eq:tracer:eom} via
second-order Lagrange interpolation. The tracers are integrated in
time via a second-order explicit Runge-Kutta scheme and, because we
integrate them in lockstep with the fluid, the coupling between
tracers and fluid is fully second-order. We utilize the particle
infrastructure already implemented in \nubhlight\ to treat tracers with
the same shared and distributed memory parallelism as for
superneutrinos.

In this work, we follow \cite{BovardTracersBNS} and roughly uniformly
sample our tracers everywhere where there is physical fluid (i.e.,
everywhere that is not atmosphere). When we set up a disk, for each
cell containing disk material, we calculate the total mass within the
cell, and equally distribute it between $N$ tracer particles that are
placed in the cell. The tracer positions within the cell are randomly
sampled from the uniform distribution.

\section{Code Verification}
\label{sec:verification}

Many aspects of \nubhlight\ have been tested rigorously in previous
works, such as \cite{HARM,grmonty,bhlight}. Here we discuss tests of
the added functionality required to study neutrino transport.

\subsection{Advection Tests}
\label{sec:ver:adv}

\begin{figure}[tpb]
  \centering
  \includegraphics[width=\columnwidth]{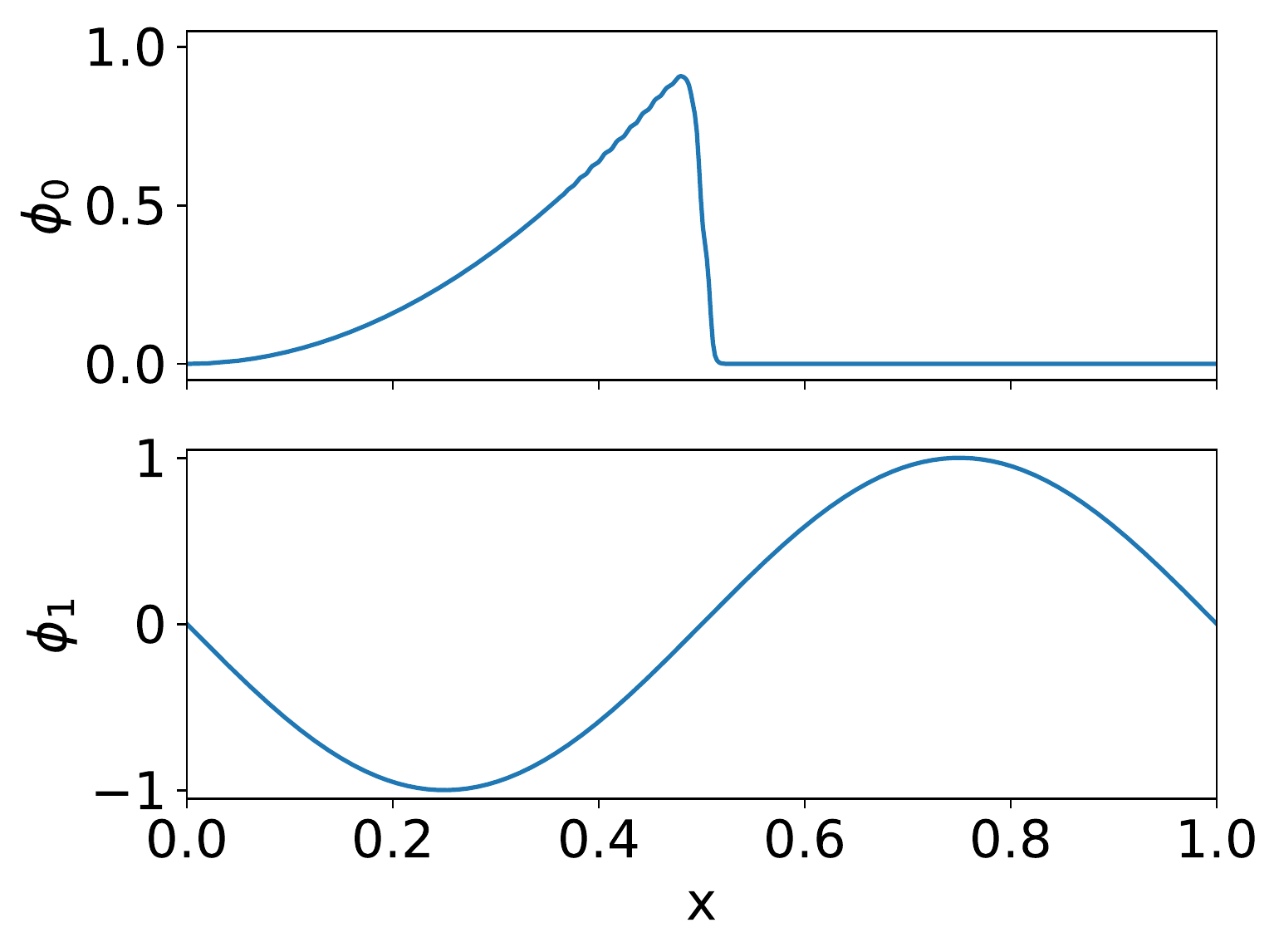}
  \caption{The solution to equation \eqref{eq:advection:test} given
    initial data \eqref{eq:adv:initial:1} and \eqref{eq:adv:initial:2}
    with periodic boundaries after one cycle.}
  \label{fig:adv:1d:x}
\end{figure}

\begin{figure}[tpb]
  \centering
  \includegraphics[width=\columnwidth]{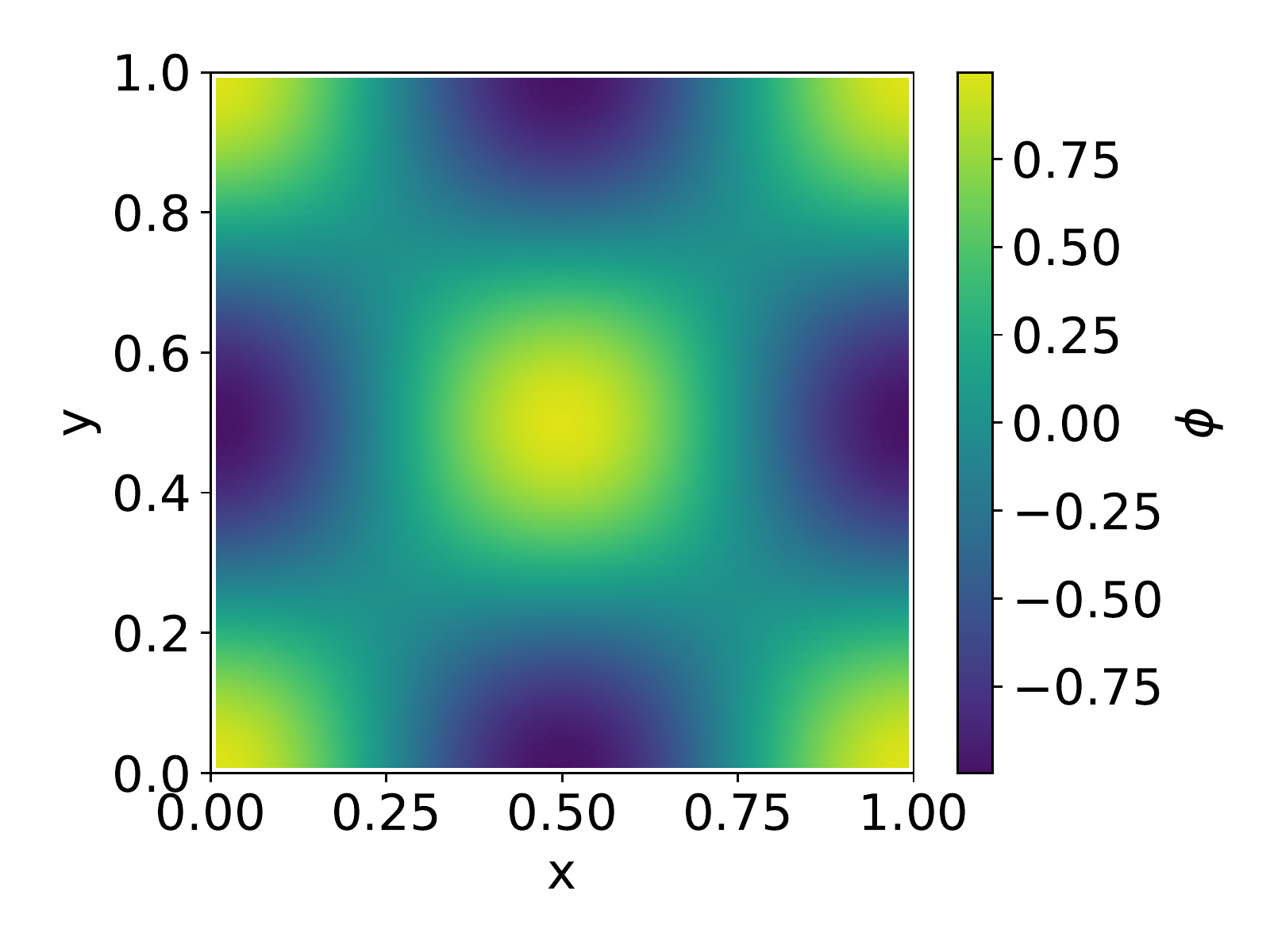}
  \caption{A two-dimensional slice of the solution to equation
    \eqref{eq:advection:test} given initial data
    \eqref{eq:adv:initial:2} in three dimensions with periodic
    boundaries after one cycle. Our grid for this calculation was
    $64\times 64\times 64$.}
  \label{fig:adv:3d:phi1}
\end{figure}

\begin{figure}[tpb]
  \centering
  \includegraphics[width=\columnwidth]{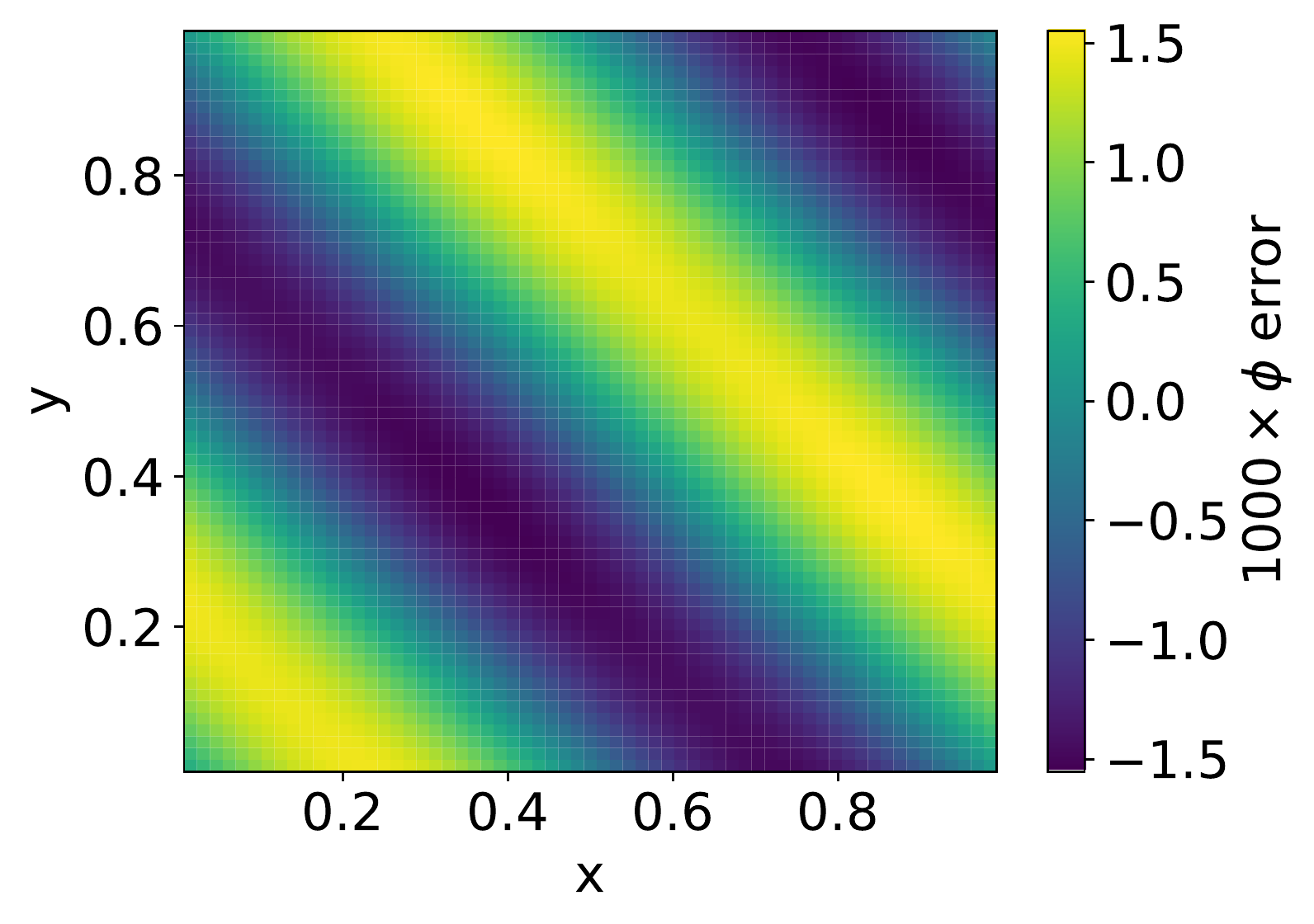}
  \caption{The pointwise error in the solution shown in figure
    \ref{fig:adv:3d:phi1}.}
  \label{fig:adv:3d:pointwise}
\end{figure}

\begin{figure}[tpb]
  \centering
  \includegraphics[width=\columnwidth]{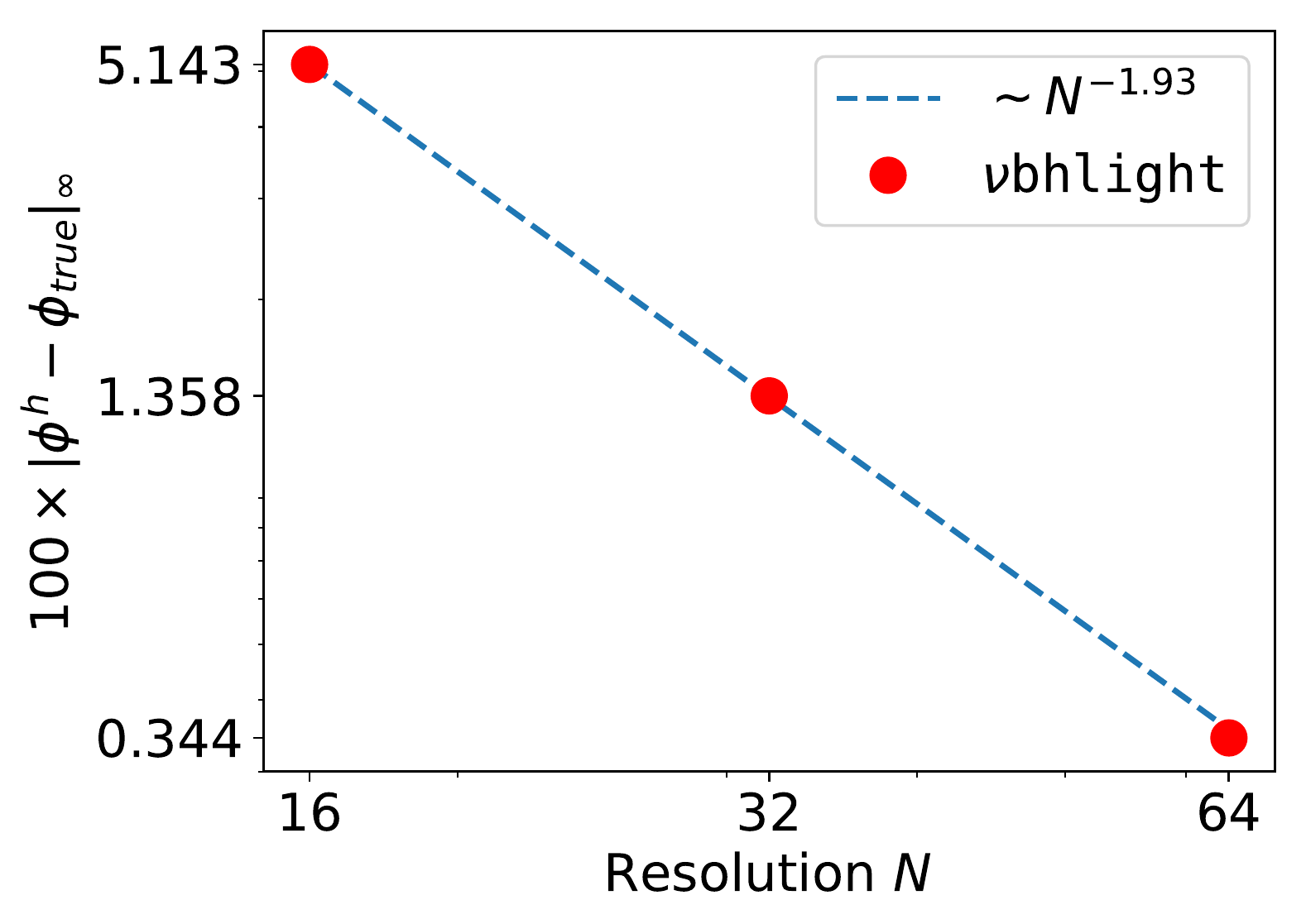}
  \caption{Convergence of the infinity norm of the error of the
    solution to equation \eqref{eq:advection:test} given initial data
    \eqref{eq:adv:initial:2} in three dimensions with periodic
    boundaries after one cycle.}
  \label{fig:adv:3d:conv}
\end{figure}

To test the passive scalar framework described in section
\ref{sec:passive:scalars}, we study the advection of an intrinsic (in
the thermodynamic sense) scalar field by a constant (in space and
time) fluid flow in flat Minkowski space. Under these conditions and
in one spatial dimension, equation \eqref{eq:passive:intrinsic}
reduces to\footnote{Although here we describe the problem setup for
  the advection of intrinsic variables, the test (and results) for
  advection of numbers is identical.}
\begin{equation}
  \label{eq:advection:test}
  \partial_t \phi  + \frac{u^x}{\sqrt{1-(u^x)^2}}\partial_x \phi = 0,
\end{equation}
where $u^x$ is the velocity of the fluid in the $x$-direction in
Minkowski coordinates.

We solve equation \eqref{eq:advection:test} as an initial value
problem using the techniques described in section
\ref{sec:meth:fluid} with initial conditions
\begin{eqnarray}
  \label{eq:adv:initial:1}
  \phi_0(t=0,x)
  &=& \begin{cases}
    4 x^2&\text{if }x \leq 0\\
    0&\text{otherwise}
  \end{cases}\\
  \label{eq:adv:initial:2}
  \text{and }\phi_1(t=0,x) &=& A\sin(2\pi x),
\end{eqnarray}
for scalar fields $\phi_0$ and $\phi_1$, both obeying equation
\eqref{eq:advection:test}, with periodic boundary conditions on the
domain $x\in [-1,1]$. We plot a solution to this initial-boundary
value problem at $t=2 \sqrt{1-(u^x)^2}/u^x$ in figure
\ref{fig:adv:1d:x}.

We use \eqref{eq:adv:initial:2} to check for convergence and
\eqref{eq:adv:initial:1} to test the handling of discontinuities. A
three-dimensional version of this test can be constructed by rotating
initial data \eqref{eq:adv:initial:2} by 45 degrees about the $y$ and
$z$ axes. We plot a two-dimensional slice of $\phi_1$ in the
three-dimensional test in figure \ref{fig:adv:3d:phi1} and a
two-dimensional slice of the pointwise error in figure
\ref{fig:adv:3d:pointwise}. Finally, we plot the convergence of
infinity norm of the error in $\phi_1$ in three dimensions in Figure
\ref{fig:adv:3d:conv}. As expected for a second-order Godunov-type method, our
solution converges at second-order.

\subsection{Tracer Particles}
\label{sec:ver:tracers}

\begin{figure}[tpb]
  \centering
  \includegraphics[width=\columnwidth]{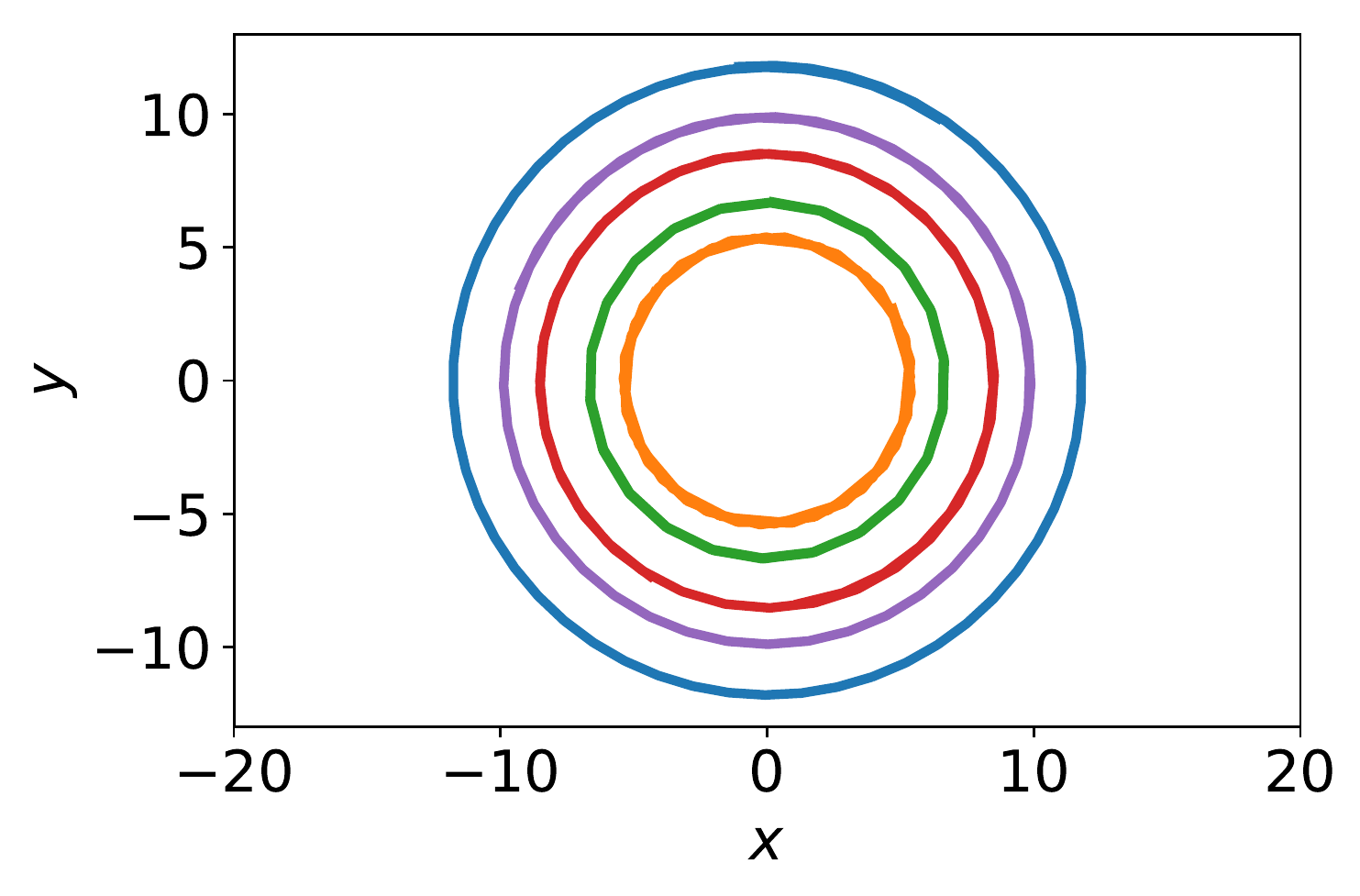}
  \caption{Tracks of tracer particles advected within an equilibrium
    torus.}
  \label{fig:tracers:3d:tracks}
\end{figure}

We test the tracer particle infrastructure described in section
\ref{sec:meth:tracers} by constructing known fluid flows and watching
the tracers advect with the fluid.

\subsubsection{Advection}

One simple known flow is that given
by equation \eqref{eq:advection:test} in section \ref{sec:ver:adv}. In
this case---where the velocity field is uniform and constant in
time---there is no truncation error in the spatial
discretization. Therefore, even for very small resolutions, errors on
the order of machine precision can easily be achieved. Indeed, we run
this test with a mere 16 cells and achieve machine precision accuracy
in the positions of the tracer particles.

\subsubsection{Equilibrium Torus}

Another simple flow is a torus in hydrostatic equilibrium about a
black hole.\footnote{The equilibrium torus is in an unstable
  equilibrium. Eventually a Papaloizou-Pringle instability will
  form. Fortunately the growth time is long compared to the simulation
  time presented here \cite{PPInstability}.} For this test, we use the
initial conditions as described in section
\ref{sec:disk:setup}. Briefly, we assume constant entropy and specific
angular momentum with a tabulated SFHo equation of state (see section
\ref{sec:disk:setup} for details). We use a relatively coarse
grid---$96\times 96\times 64$ cells and $\sim 657,000$ tracer
particles. We evolve the system for $500$ gravitational times
($G M_{BH}/c^3$) with no seed magnetic field.

The continuum initial data is in equilibrium, but the numerical
initial data is not. We allow the disk to relax towards numerical
equilibrium for $200$ gravitational times, then select tracer
particles within $1 \times GM_{BH}/c^2$ of the midplane of the
disk. Figure \ref{fig:tracers:3d:tracks} shows projections of tracks
of a random selection of these tracer particles onto the
$xy$-plane. The dynamical time in the inner region is shorter than in
the outer region, so the innermost trace covers many orbits, while the
outermost trace covers only one. Similarly, the outermost region is
not yet in numerical equilibrium, hence why the track does not
close.\footnote{Since integration is not symplectic, we do not expect
  orbits to completely close, even with a perfectly relaxed
  disk. However, compared to the effect of disk relaxation, this
  effect is negligible and it is not visible here.}

\subsection{Fake Table Tests}
\label{sec:ver:fake}

To test our tabulated EOS reader, we tabulate the ideal gas law 
\begin{equation}
  \label{eq:ideal:gas:law}
  P = (\Gamma - 1) u,
\end{equation}
where $\Gamma$ is the ratio of specific heats. Using this ``fake''
table, we can repeat tests presented in \cite{bhlight} in the absence
of radiation. We perform the non-relativistic linear waves and shock
tube tests presented in \cite{bhlight}.

The tables treat all quantities on a logarithmic scale. For equation
\eqref{eq:ideal:gas:law}, the log of all thermodynamic quantities
except sound speed is linear and the interpolation is exact, yielding
identical results to an analytic EOS. The logarithm of the sound speed
is linear at low velocities, but not at relativistic
velocities. However, the tests we reproduce from \cite{bhlight}
are non-relativistic and so these nonlinearities are not
present. Therefore, we expect agreement up to machine precision. And
indeed, we find this to be the case.

\subsection{Artificial Neutrino Cooling}
\label{sec:ver:nu}

\begin{figure}[tpb]
  \centering
  \includegraphics[width=\columnwidth]{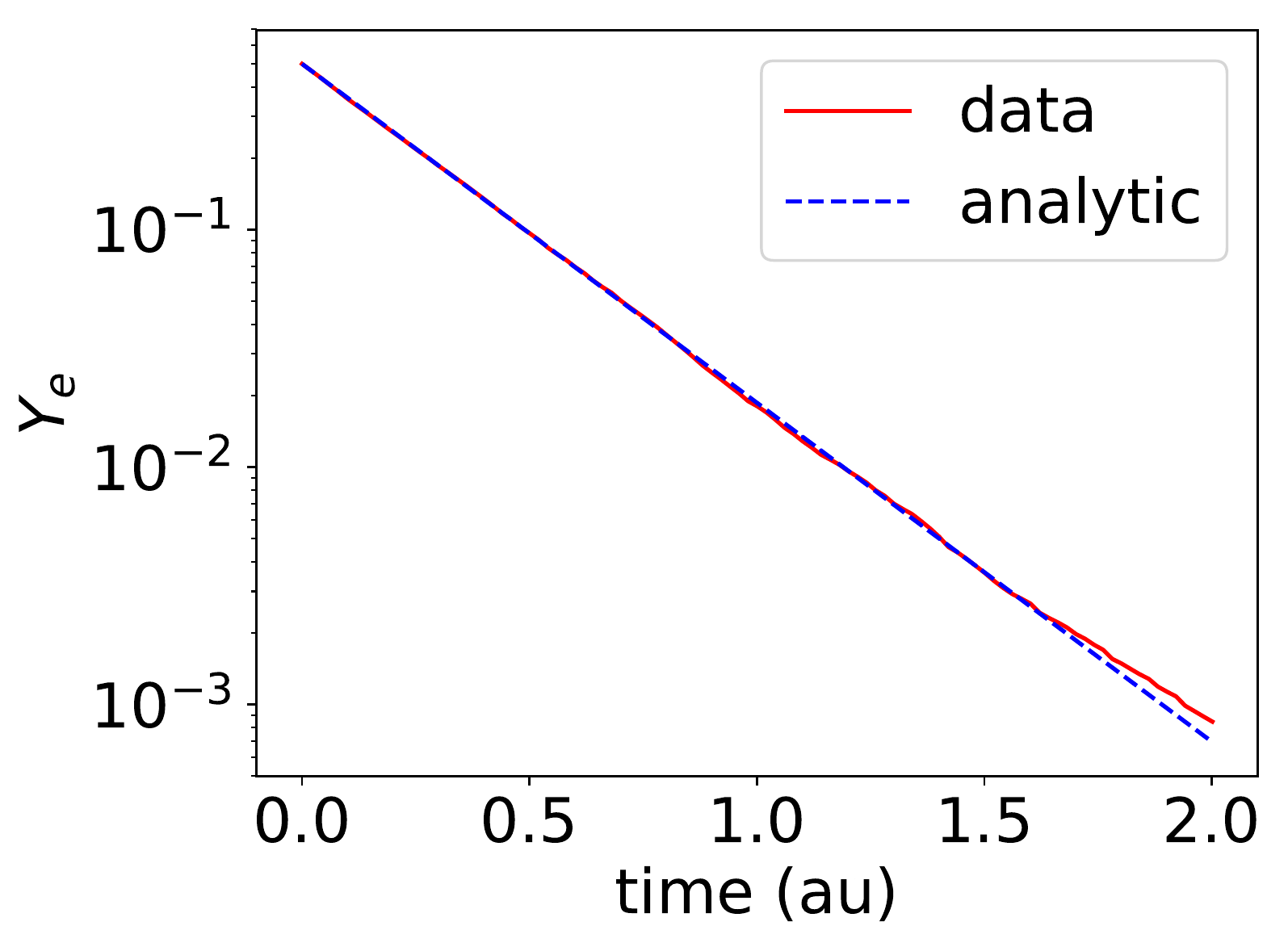}
  \caption{The electron fraction $Y_e$ as a function of time for a
    homogeneous isotropic gas cooled by electron neutrinos. The solid
    line is the analytic solution, and the dashed line is the measured
    data. Agreement is very good.}
  \label{fig:cool:nu:ye}
\end{figure}

\begin{figure}[tpb]
  \centering
  \includegraphics[width=\columnwidth]{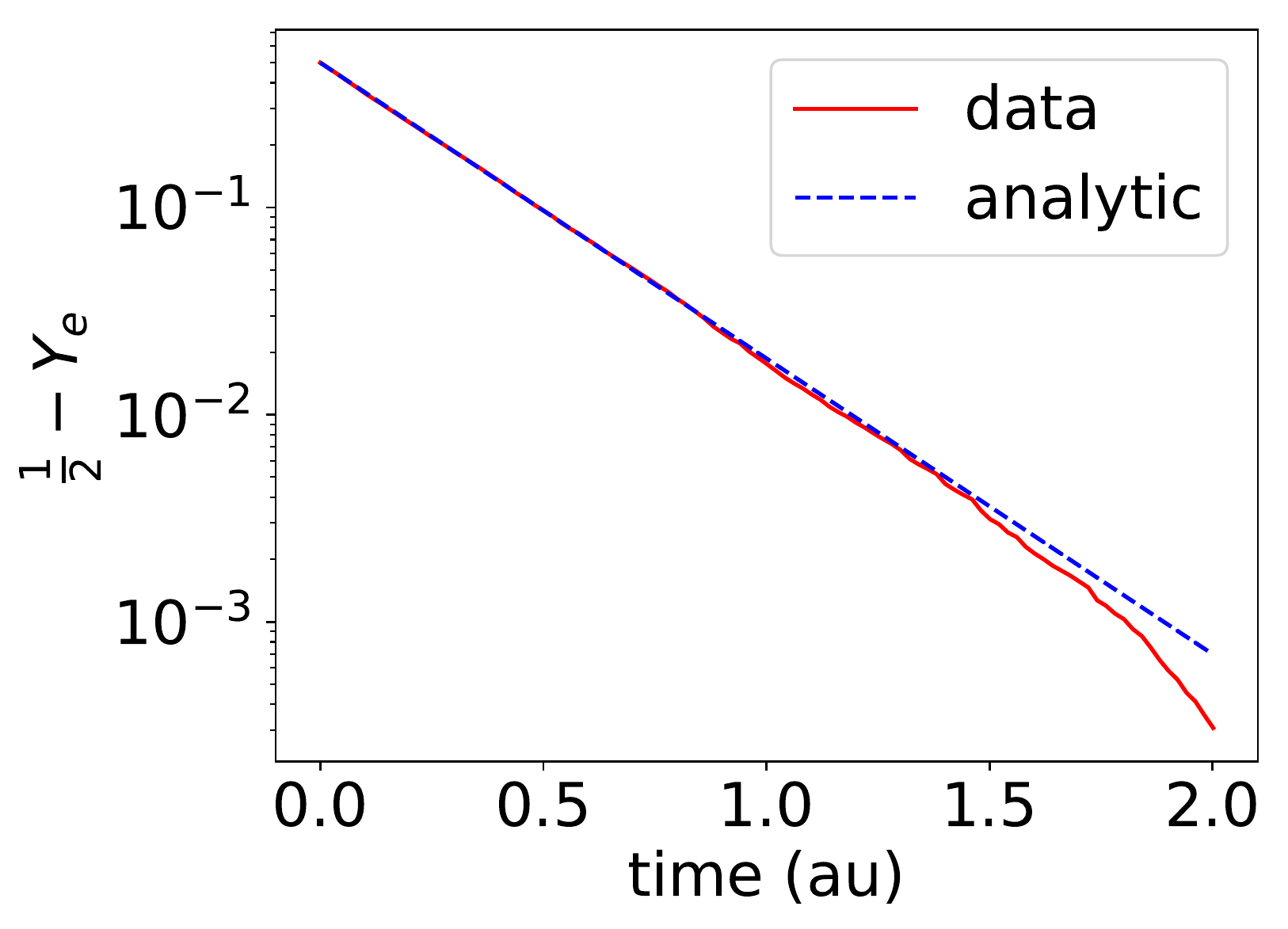}
  \caption{The electron fraction $Y_e$ as a function of time for a
    homogeneous isotropic gas cooled by electron antineutrinos. The solid
    line is the analytic solution, and the dashed line is the measured
    data. Agreement is very good.}
  \label{fig:cool:antinu:ye}
\end{figure}

\begin{figure}[tpb]
  \centering
  \includegraphics[width=\columnwidth]{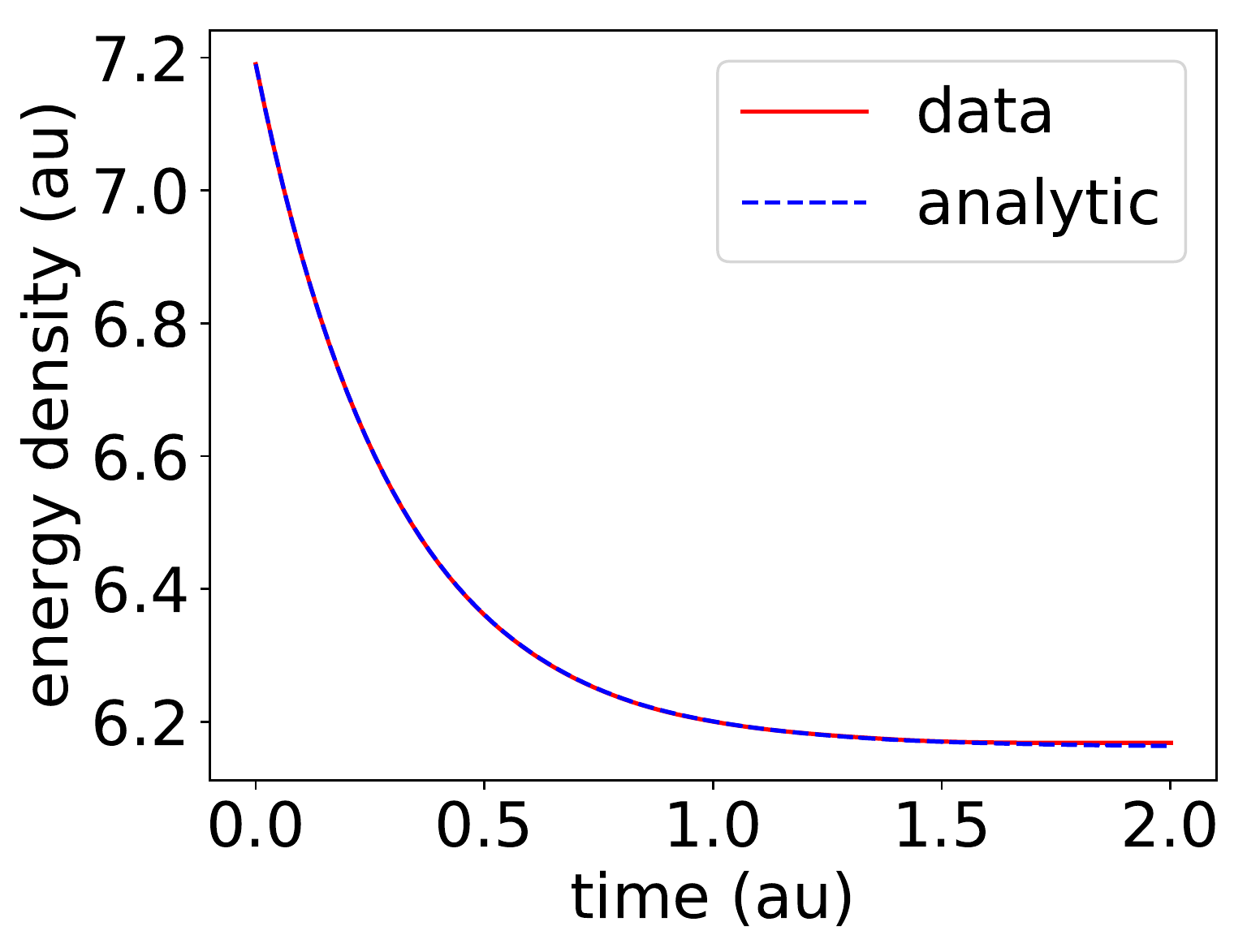}
  \caption{The energy density $u$ as a function of time for a
    homogeneous isotropic gas cooled by neutrinos.  The solid
    line is the analytic solution, and the dashed line is the measured
    data. Agreement is very good.}
  \label{fig:cool:nu:u}
\end{figure}

As a basic test of the coupling of neutrinos to matter, we study
optically thin neutrino cooling in a simplified context. For this
test, we choose to emit either only electron neutrinos or only
electron antineutrinos. In each case, we define an emissivity of the
form
\begin{equation}
  \label{eq:ver:cool:jnu}
  \jnuf = C y_f(Y_e) \chi([\numin,\numax])
\end{equation}
where $C$ ensures the units and scale are appropriate,
\begin{equation}
  \label{eq:ver:cool:chi}
  \chi([\numin,\numax]) =
  \begin{cases}
    1&\text{if }\numin\leq\nu\leq\numax\\
    0&\text{otherwise}
  \end{cases}
\end{equation}
is the selection function, and $y_f(Y_e)$ is given by
\begin{equation}
  \label{eq:ver:cool:yf}
  y_f(Y_e) =
  \begin{cases}
    2 Y_e&\text{if emitting } \nu_e\\
    1-2Y_e&\text{if emitting } \bar{\nu}_e\\
    0&\text{otherwise}
  \end{cases}.
\end{equation}

Assuming a homogeneous and isotropic fluid at rest, this implies that
the electron fraction $Y_e$ and internal energy density $u$ obey
ordinary differential equations of the form
\begin{eqnarray}
  \label{eq:ver:cool:dye}
  \partial_t Y_e &=& - A_C y_f(Y_e)\\
  \label{eq:ver:cool:dtu}
  \text{and }\partial_t u &=& -B_C y_f(Y_e),
\end{eqnarray}
where
\begin{eqnarray}
  \label{eq:ver:cool:def:AC}
  A_C &=& \frac{m_p}{h\rho}C\ln\paren{\frac{\numax}{\numin}}\\
  \label{eq:ver:cool:def:BC}
  \text{and } B_c &=& C \paren{\numax - \numin}.
\end{eqnarray}
Equation \eqref{eq:ver:cool:dye} has a solution
\begin{equation}
  \label{eq:eq:var:cool:ye:of:t}
  Y_e(t) =
  \begin{cases}0\\\frac{1}{2}\end{cases}
  +
  e^{-2 A_C t}
  \begin{cases}
    Y_e(t=0) &\text{for }\nu_e\\
    \sqrbrace{(Y_e)_0-\frac{1}{2}}&\text{for }\bar{\nu}_e
  \end{cases},
\end{equation}
where $(Y_e)_0 = Y_e(t=0)$. This implies that the electron fraction
either exponentially decays to zero or exponentially approaches $1/2$,
depending on whether we emit electron neutrinos or electron
antineutrinos. With this solution in hand, we can solve equation
\eqref{eq:ver:cool:dtu} to find that
\begin{equation}
 \label{eq:var:cool:u:of:t}
 u(t) = u_0 + \frac{B_C}{A_C}\paren{e^{-2 A_c t}-1}
 \begin{cases}
   (Y_e)_0&\text{for }\nu_e \\
   \frac{1}{2} - (Y_e)_0 &\text{for }\bar{\nu}_e
 \end{cases},
\end{equation}
where $u_0 = u(t=0)$.

In our tests, we choose
\begin{equation}
  \label{eq:cool:Ye0}
  Y_e(t=0) =
  \begin{cases}
    \frac{1}{2}&\text{for }\nu_e\\
    0&\text{for }\bar{\nu}_e
  \end{cases}
\end{equation}
so that equation \eqref{eq:eq:var:cool:ye:of:t} reduces to
\begin{displaymath}
  Y_e(t) =
  \begin{cases}
    -\frac{1}{2}e^{-2A_Ct}&\text{for }\nu_e\\
    \frac{1}{2}\paren{1 - e^{-2A_Ct}}&\text{for }\bar{\nu}_e
  \end{cases}
\end{displaymath}
and equation \eqref{eq:var:cool:u:of:t} reduces to
\begin{displaymath}
  u(t) = u_0 + \frac{B_C}{2A_C}\paren{e^{-2 A_c t}-1}
\end{displaymath}
and $u(t)$ asymptotes to $u_0 - B_C/(2A_C)$. Figure
\ref{fig:cool:nu:ye} shows the electron fraction as a function of time
for a gas cooled by electron neutrinos and figure
\ref{fig:cool:antinu:ye} shows the analogous quantity for a gas cooled
by electron antineutrinos. The energy density as a function of time
looks identical whether we cool by electron neutrinos or electron
antineutrinos. We plot this in figure \ref{fig:cool:nu:u}. In all
cases, the agreement is good. A small deviation appears at late times,
but since the plot is on a log scale, this deviation is extremely
small.

\subsection{Artificial Single Scattering Events}
\label{sec:ver:scatt}

\begin{figure}[tpb]
  \centering
  \includegraphics[width=\columnwidth]{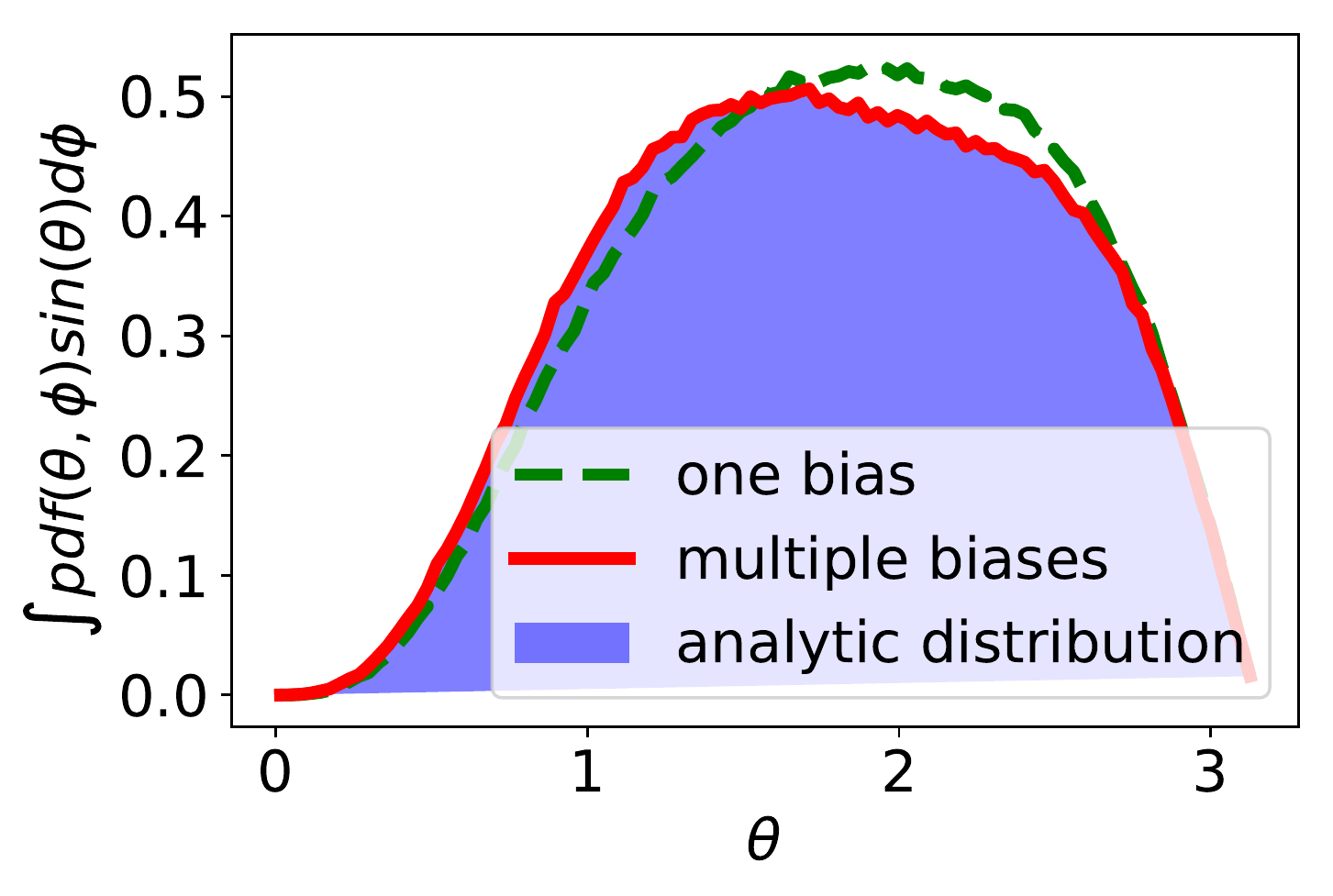}
  \caption{The distribution of scattered physical neutrinos as a
    function of angle $\theta$. The blue shaded region represents the
    integral of the analytic solution. The green dashed line is
    measured data with a single global bias parameter. The red solid
    line is measured data with multiple per-scattering-process
    biases. Scattering with per-process biases matches the analytic
    solution very well, while scattering with a single global bias
    undersamples some processes and oversamples others.}
  \label{fig:multiscatt}
\end{figure}

Our procedure for biasing scattering on a per-interaction basis is a
novel part of \nubhlight. Therefore it is worth checking both that it
works well and that it is worth the added complexity. We seek to test
this here. In this test, we take a homogeneous, zero temperature gas
and propagate a steady stream of heavy neutrinos traveling in the
$z$-direction through it so that each neutrino traverses a scattering
optical depth for the most likely scattering interaction of roughly
$\Delta \tau = 1$. The optical depth for the less likely scattering
interactions will be less than unity, the least likely interaction
significantly so. If the resolution is chosen so that the most likely
scattering interaction is just barely well-sampled, a naive biasing
algorithm will undersample these less likely interactions.

After propagating the neutrinos, we investigate the
directions of the neutrinos that have scattered exactly once. The
probability distribution of directions traveled by these scattered
neutrinos should match the total probability distribution formed by
the sum of all differential scattering cross sections
\begin{equation}
  \label{eq:test:multiscatt:sigmasum}
  \frac{dN}{d\Omega} \sim \sum_{p} \frac{d\sigma_{p}}{d\Omega}
\end{equation}
where $N(\theta,\phi)$ is the total number of heavy neutrinos
traveling in the $(\theta,\phi)$ direction and $d\sigma_p/d\Omega$ is
the differential cross section for a heavy neutrino scattering off of
a gas particle of species $p$. We measure $N$ integrated over the
azimuthal direction:
\begin{equation}
  \label{eq:test:multiscatt:sigmasum:intphi}
  \frac{dN}{d\theta}(\theta) \sim \sum_p \int d\phi \sin(\theta)\frac{d\sigma_p}{d\theta}.
\end{equation}
For the purpose of this test, we introduce three fake particles, with
three fake, anisotropic, elastic scattering kernels:
\begin{equation}
  \label{eq:test:multiscatt:dsigma:domega}
  \frac{d\sigma_i}{d\Omega} = \sigma_0 (4i+1)(1+\mu^{2i+1}),\ i=0,1,2,
\end{equation}
where $\mu=\cos(\theta)$. These interactions have different total
cross sections and thus different probabilities that an individual
neutrino will scatter via a given process. However, if all processes
are well sampled, we should be able to measure a probability
distribution that matches equation
\eqref{eq:test:multiscatt:sigmasum:intphi}.

We perform this experiment in two ways. First, we use a single global
bias which modifies scattering probability uniformly across all
interactions. For a resolution which marginally well-samples the most
likely interaction ($i=2$), the less likely interactions ($i=0,1$)
will be undersampled. Second, we bias each scattering interaction
individually, as described in section \ref{sec:meth:rad:scatt}. This
second approach should more evenly sample all interactions for a given
resolution. In both cases, we use the same number of superneutrinos
(roughly $10^6$) and set the biases such that the same number of
unscattered superneutrinos scatters each timestep (roughly
$4\times 10^4$).

We show our results in figure \ref{fig:multiscatt}. The area of the
blue shaded region is the integral $\int (dN/d\Omega) d\Omega$ of
equation \eqref{eq:test:multiscatt:sigmasum}, meaning the boundary of
the shaded region is given by equation
\eqref{eq:test:multiscatt:dsigma:domega}. The green dashed curve is
the probability distribution of superneutrinos measured when the
experiment is performed with a single global bias. The red solid curve
is the probability distribution measured when the experiment is
performed with per-interaction biases.

When global biases are used $d\sigma_0/d\Omega$ and
$d\sigma_1/d\Omega$ are undersampled with respect to
$d\sigma_2/d\Omega$. But when per-interaction biases are used, the
agreement with equation \eqref{eq:test:multiscatt:sigmasum:intphi} is
quite good. This indicates both the necessity and efficacy of
per-interaction biases.

\subsection{Two-Dimensional Lepton Transport}
\label{sec:test:ye2d}

\begin{figure*}[t!]
  \centering
  \includegraphics[width=\textwidth]{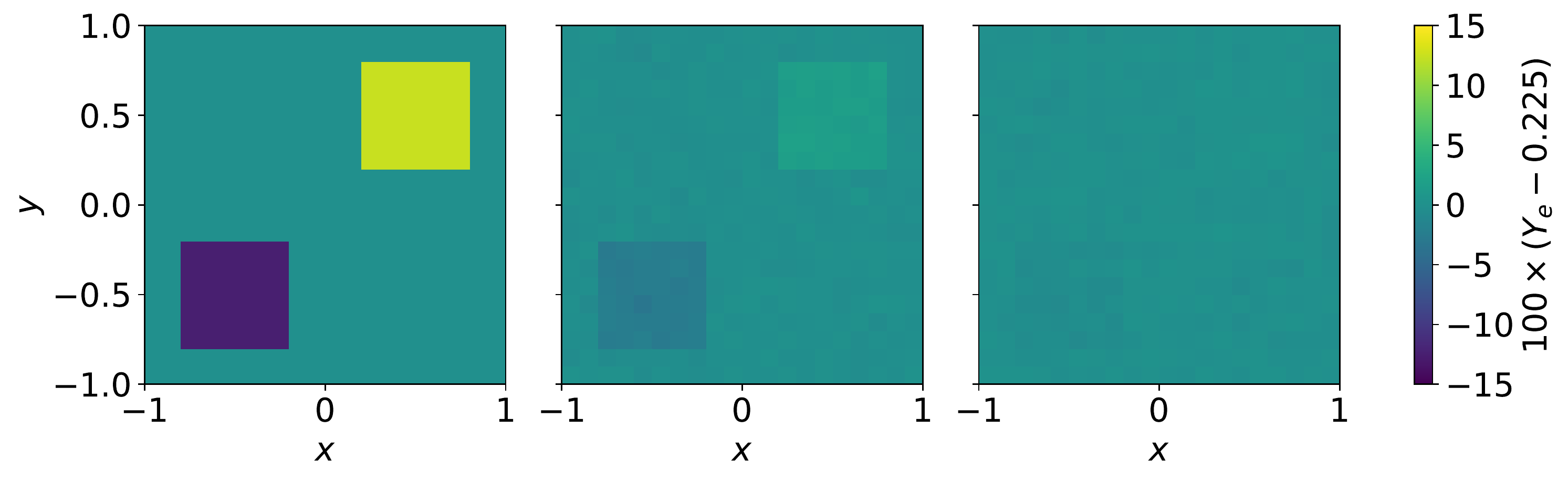}
  \caption{Neutrinos equilibriate electron fraction between a hot spot
    and a cold spot as time passes. Left: the system at the initial
    time. Center: The system after $\sim 5$ ms. Right: The system
    after $\sim 10$ ms.}
  \label{fig:test:ye2d:frames}
\end{figure*}

\begin{figure}
  \centering
  \includegraphics[width=\columnwidth]{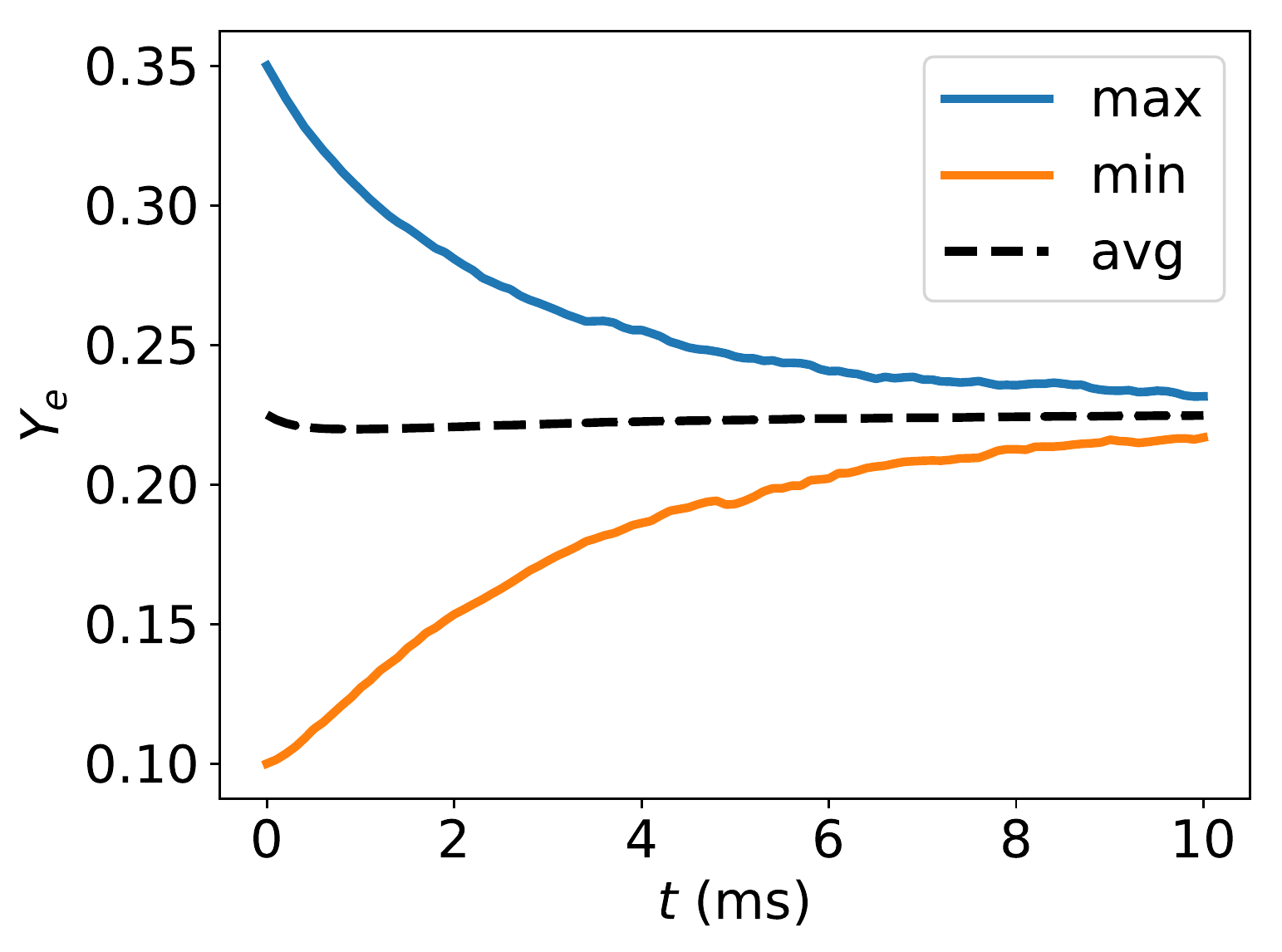}
  \caption{Electron fraction as a function of time for the hot spot
    (blue), the cold spot (orange), and the remainder of the gas
    (black). The average electron fraction experiences an early
    transient as leptons are carried into the radiation field but then
    remains stable. The hot spot and cold spot converge to the average
    exponentially with time.}
  \label{fig:test:ye2d:time}
\end{figure}

A major motivation for treating neutrino radiation accurately is the
fact that neutrinos can carry lepton number from one place to
another. In the context of an accretion disk, this means leptons---and
thus electron fraction---can be transported from one part of the disk
to another.

To demonstrate this capability in \nubhlight, we perform the following
simple two-dimensional test. Consider a gas in a two-dimensional,
periodic box,
\begin{equation}
  \label{eq:2dye:dom}
  (x,y) \in [-1,1]^2.
\end{equation}
The gas is at constant density and
temperature
\begin{eqnarray}
  \label{eq:2dye:rho}
  \rho &=& 10^{10}\ \text{g}/\text{cm}^3\\
  \label{eq:2dye:T}
  T    &=& 2.5\ \text{MeV},
\end{eqnarray}
and piecewise-constant electron fraction defined by
\begin{equation}
  \label{eq:2dye:Ye}
  Y_e =
  \begin{cases}
    0.1   &\text{if }(x,y)\in [-0.75,-0.25]^2\\
    0.35  &\text{if }(x,y)\in [0.25,0.75]^2\\
    0.225 &\text{otherwise}
  \end{cases},
\end{equation}
so that there is one ``hot spot'' of $Y_e$ and one ``cold spot.'' The
hot and cold spot regions are separated from the rest of the gas by
membranes which are impermeable by the gas but through which neutrinos
can travel freely. In this way, the gas does not evolve due to
pressure gradients.

Over time, as neutrinos are emitted and absorbed, the electron
fractions in the hot spot and the cold spot will come to equilibrium
with each other. Indeed, the gas itself will come into equilibrium
with the radiation field. Figure \ref{fig:test:ye2d:frames} shows the
electron fraction as a function of space for three times. Figure
\ref{fig:test:ye2d:time} shows the evolution of the electron fraction
as a function of time. The average electron fraction experiences an
early transient as leptons are carried into the radiation field but
then remains stable. The hot spot and cold spot converge to the
average exponentially with time. The final electron fraction is not
the average $Y_e$ in the initial condition.

\subsection{Code Comparisons}
\label{sec:test:fornax}

\begin{figure}[tpb]
  \centering
  \includegraphics[width=\columnwidth]{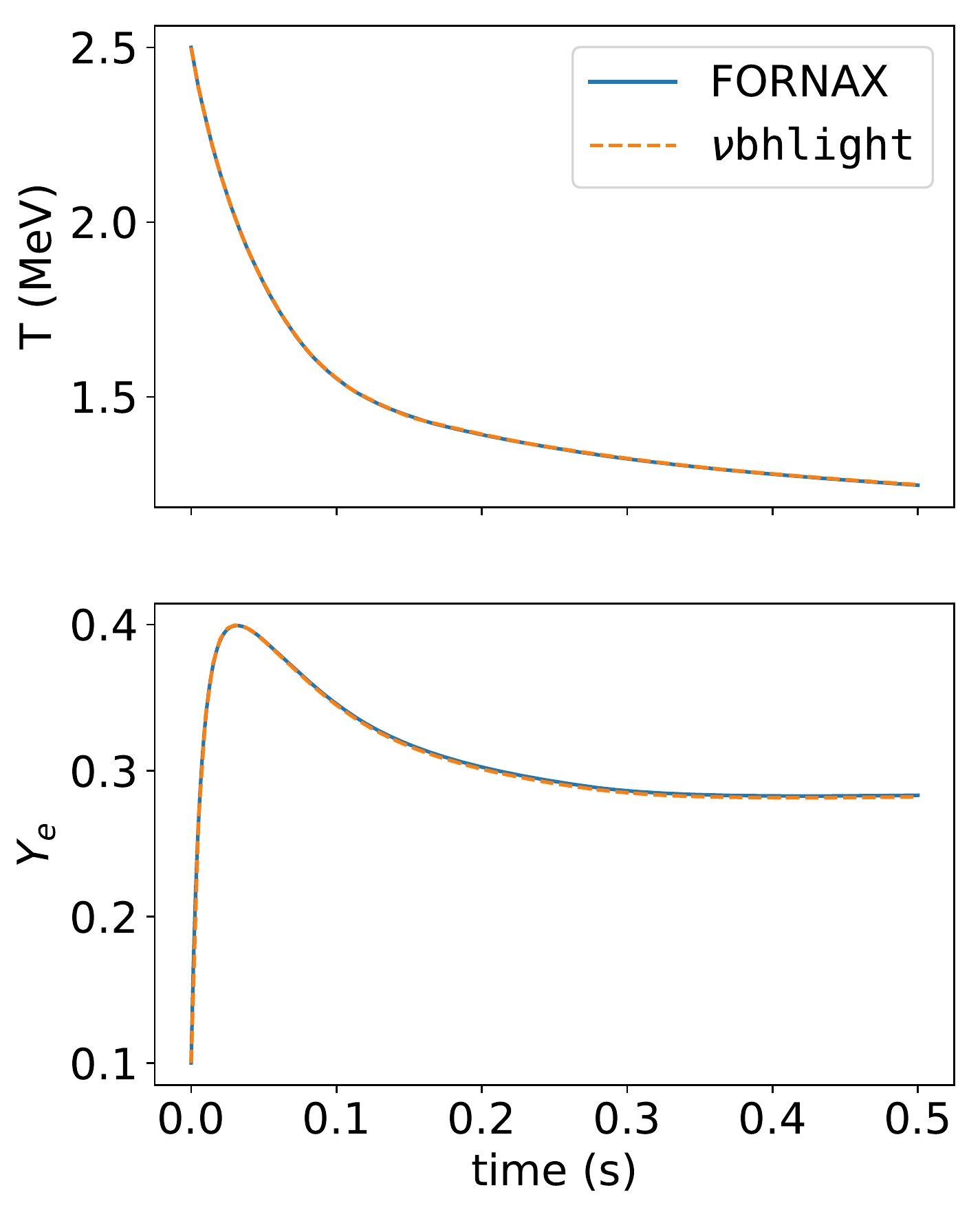}
  \caption{Temperature $T$ and electron fraction $Y_e$ for the
    optically thin cooling comparison between \nubhlight${}$ and
    \fornax. The electron fraction rapidly grows as the gas
    cools.}
  \label{fig:fornax:cooling}
\end{figure}

\begin{figure}[tpb]
  \centering
  \includegraphics[width=\columnwidth]{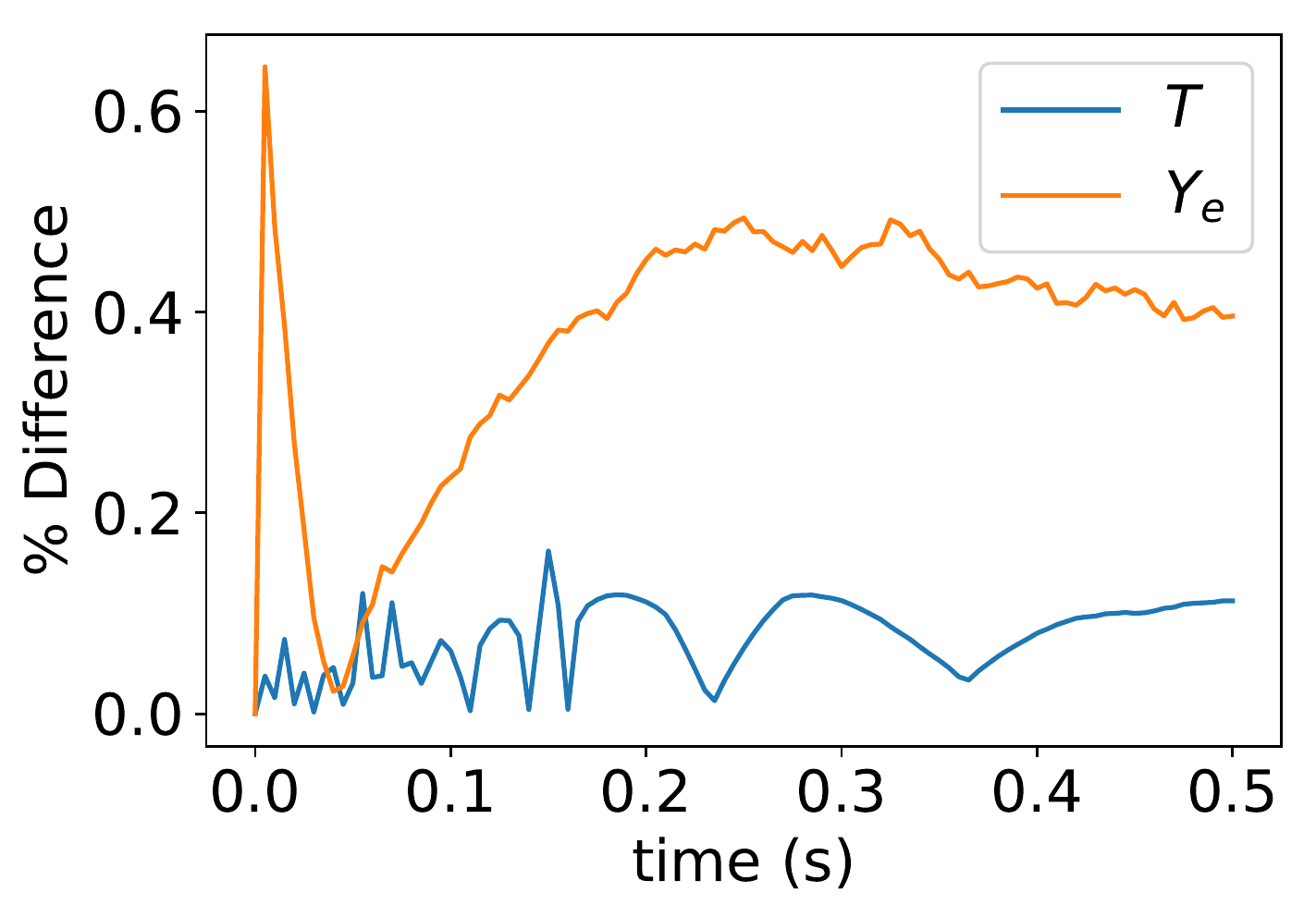}
  \caption{Percent difference in Temperature $T$ and electron fraction
    $Y_e$ for the optically thin cooling comparison between \nubhlight${}$
    and \fornax. }
  \label{fig:fornax:cooling:diff}
\end{figure}

\begin{figure}[tpb]
  \centering
  \includegraphics[width=\columnwidth]{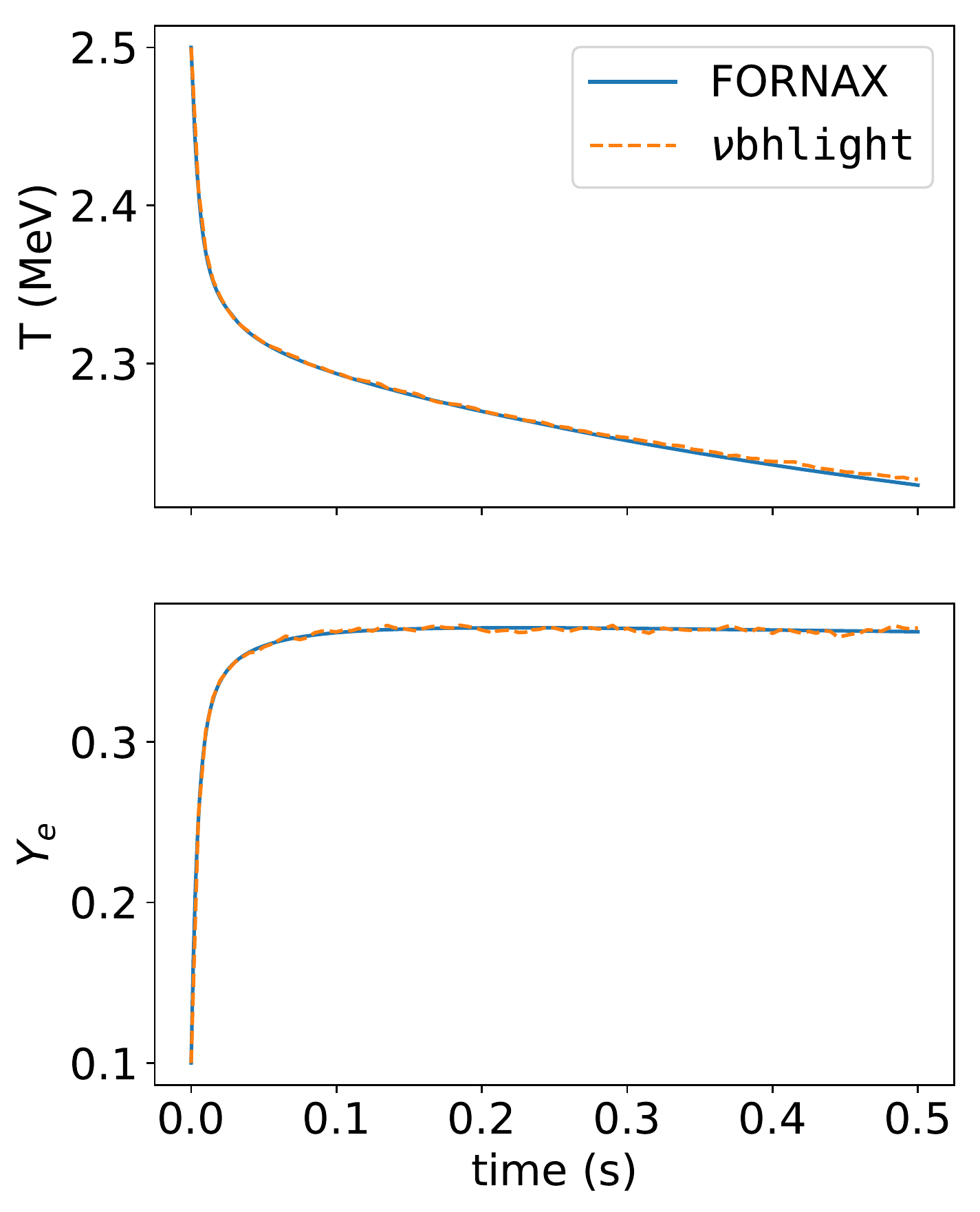}
  \caption{Temperature $T$ and electron fraction $Y_e$ for the
    thermal equilibrium comparison between \nubhlight${}$ and
    \fornax. The electron fraction rapidly grows as the gas cools
    before reaching equilibrium.}
  \label{fig:fornax:equil}
\end{figure}

\begin{figure}[tpb]
  \centering
  \includegraphics[width=\columnwidth]{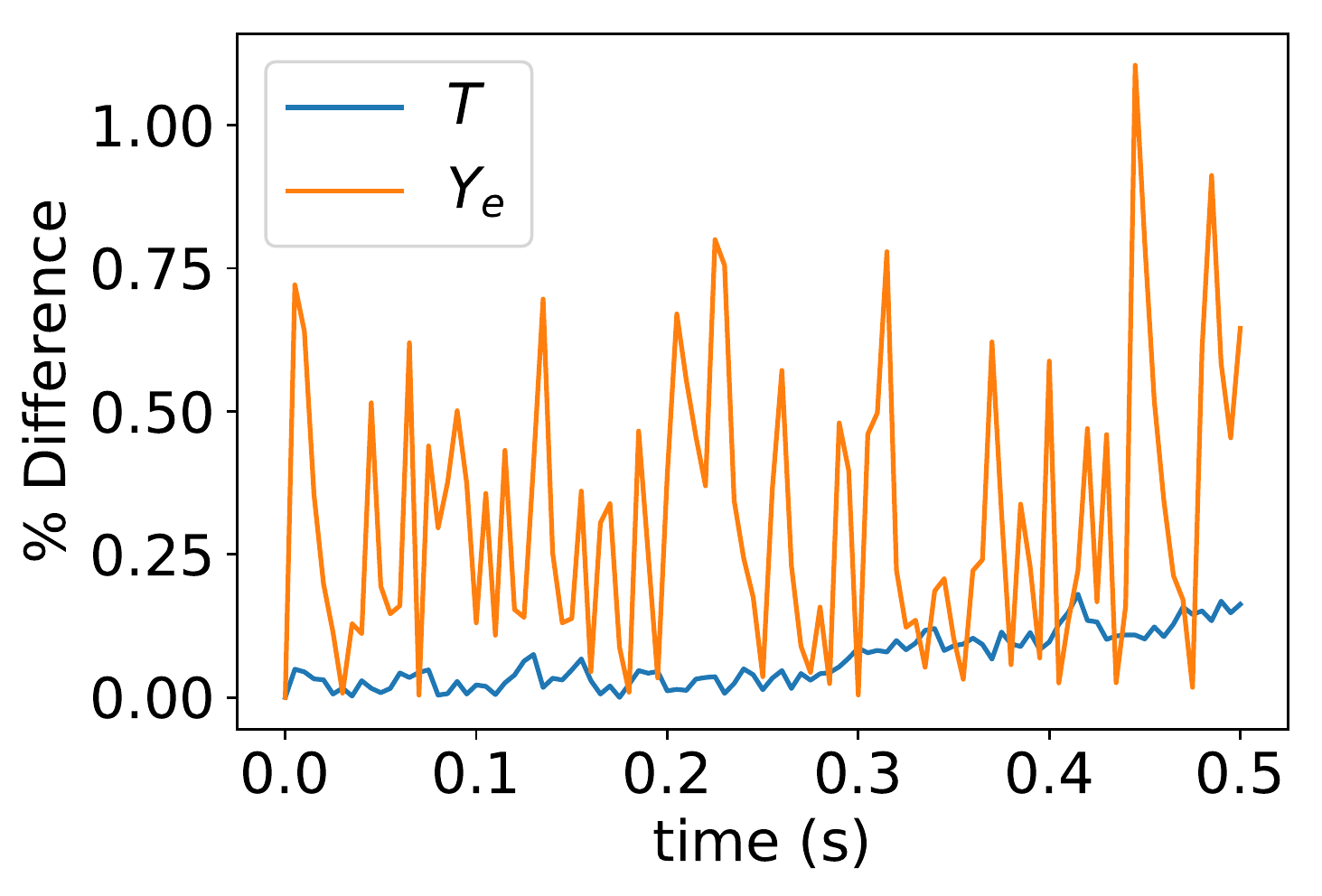}
  \caption{Percent difference in Temperature $T$ and electron fraction
    $Y_e$ for the thermal comparison between \nubhlight${}$
    and \fornax. }
  \label{fig:fornax:equil:diff}
\end{figure}

The tests described in sections \ref{sec:ver:nu} and
\ref{sec:ver:scatt} use artificial emissivities and scattering cross
sections. This has the advantage of permitting an analytic solution
against which we can compare. However, it has the significant
disadvantage of being unphysical. We would also like to test
the performance of our physical emissivities and absorption opacities.

To this end, we compare our code to the supernova code \fornax\
\cite{fornax} in two simple test cases. The two codes are designed for
different scenarios and use significantly different
methods. \nubhlight${}$ is fully general relativistic, whereas \fornax\
uses non-relativistic dynamics and an approximate treatment for
gravity. \nubhlight${}$ uses Monte Carlo transport for neutrinos, while
\fornax\ uses a multi-group moment formalism with the M1 closure model
\cite{castor2004radiation}. Some care is thus required to choose test
cases where both codes converge to the same, physically correct,
solution.

We therefore use a simple zero-dimensional setup. We use a homogenous
and isotropic gas at rest on a periodic domain in Minkowski space. We
also use the same equation of state---SFHo by Steiner et
al. \cite{SFHoEOS}. This eliminates discrepancies due to treatment of
the gas. We use the same emission and absorption opacities, as
presented in \cite{BurrowsNeutrinos,fornax} and provided by
\cite{BurrowsCorresp}. Since the scattering cross-sections are
different between the codes, we disable scattering for these
tests. Moreover, by studying only homogeneous, isotropic radiation, we
enter a regime where the M1 closure model is valid.

In both our comparison tests, we use the following initial conditions
for the gas:
\begin{eqnarray}
  \label{eq:fornax:id:rho}
  \rho_0(t = 0) &=& 10^9\ \text{g}/\text{cm}^3\\
  \label{eq:fornax:id:T}
  T(t = 0) &=& 2.5\ \text{MeV}\\
  \label{eq:fornax:id:Ye}
  Y_e(t = 0) &=& 0.1
\end{eqnarray}
which roughly mimic conditions one might encounter in a
neutrino-driven accretion flow. We run each calculation for total
duration of 0.5 seconds. We assume that there is no radiation at the
initial time. In both cases we run \fornax${}$ with 200 energy groups and
energies ranging from 1 to 300 MeV. We run \nubhlight${}$ with a target
number of $10^5$ superneutrinos which can have energies in the same
range as in \fornax.

\subsubsection{Optically Thin Cooling}
\label{sec:test:fornax:optically:thin:cooling}

In this test, we enable emissivity but disable absorption and
scattering. Traces of electron fraction and temperature are shown for
both \fornax${}$ and \nubhlight${}$ in figure \ref{fig:fornax:cooling}. The
cooling rate is a steep function of temperature. As the gas cools
rapidly, the electron fraction changes rapidly before reaching
equilibrium.

The agreement between \fornax and \nubhlight is at the percent level,
as shown in figure \ref{fig:fornax:cooling:diff}. Given that the codes
use dramatically different methods, this agreement is quite
satisfactory for the problem of interest.

\subsubsection{Thermalization}
\label{sec:test:fornax:thermalization}

In this test, we enable both emission and absorption but disable
scattering. The goal is to watch as the gas and the radiation reach
thermal equilibrium. We plot the electron fraction and the temperature
for both \fornax\ and \nubhlight\ in figure
\ref{fig:fornax:equil}. The electron fraction changes very rapidly,
but the cooling rate is subdued thanks to absorption.

\nubhlight\ and \fornax\ again agree within about a percent, as shown in
figure \ref{fig:fornax:equil:diff}. Again, given that the codes use
dramatically different methods, this agreement is quite good.

\section{Post-Merger Disk}
\label{sec:disk}

As a demonstration of \nubhlight's capabilities, we perform a fully
three-dimensional neutrino radiation GRMHD simulation of a
representative accretion disk that formed from a compact binary
merger.
Although we do not perform a detailed analysis, we believe our
qualitative results compellingly demonstrate both the capabilities of
our code and a need for those capabilities.

\subsubsection{Disk Setup}
\label{sec:disk:setup}

\begin{figure}[tpb]
  \centering
  \includegraphics[width=\columnwidth]{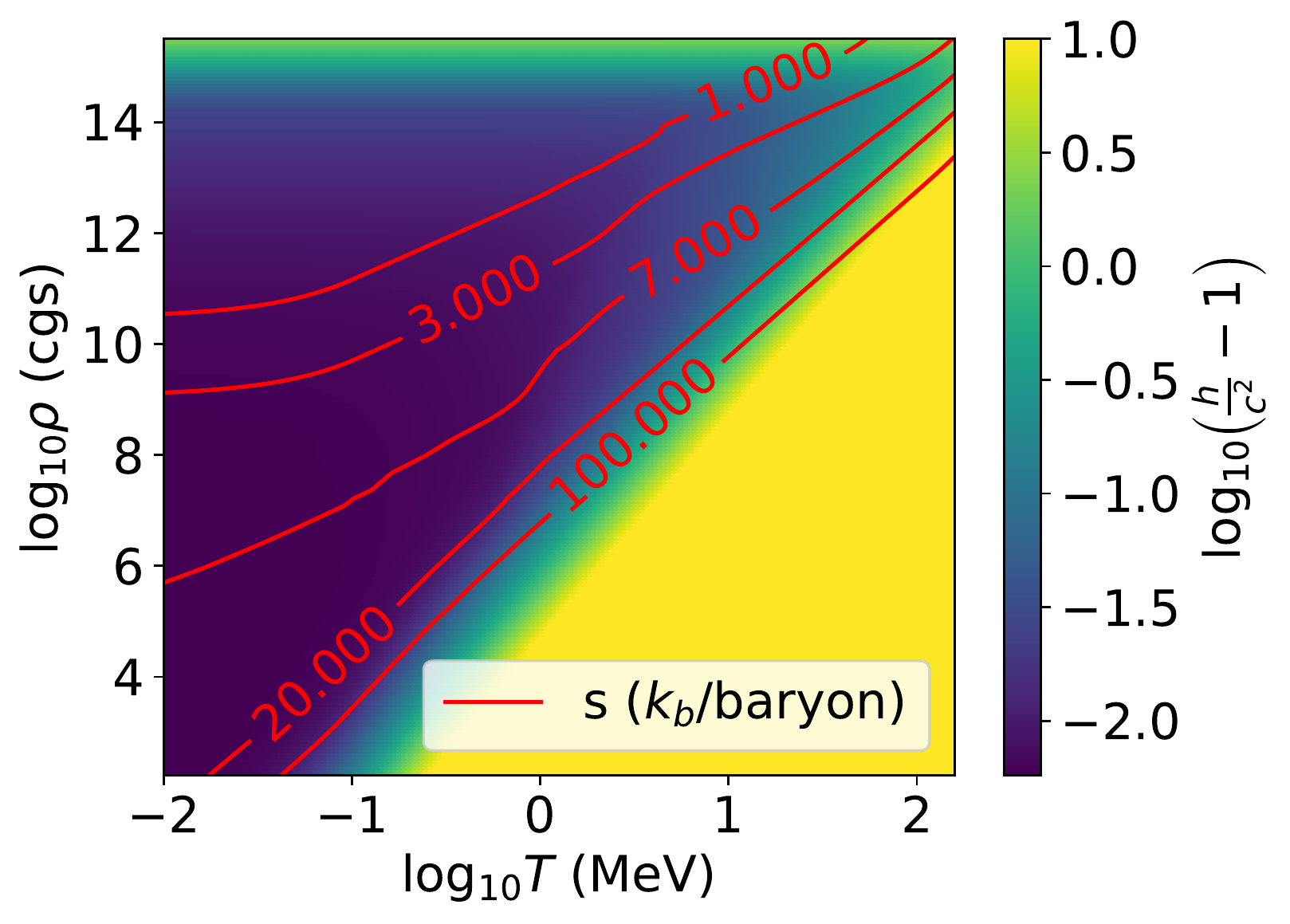}
  \caption{Contours of constant entropy (in units of $k_b/$baryon)
    superposed on a plot of specific enthalpy in terms of $\lrho$ and
    $\lT$ for the Hempel DD2 equation of state. Here we assume a
    constant electron fraction of $Y_e=0.1$.}
  \label{fig:s:contours}
\end{figure}

\begin{figure}[tpb]
  \centering
  \includegraphics[width=\columnwidth]{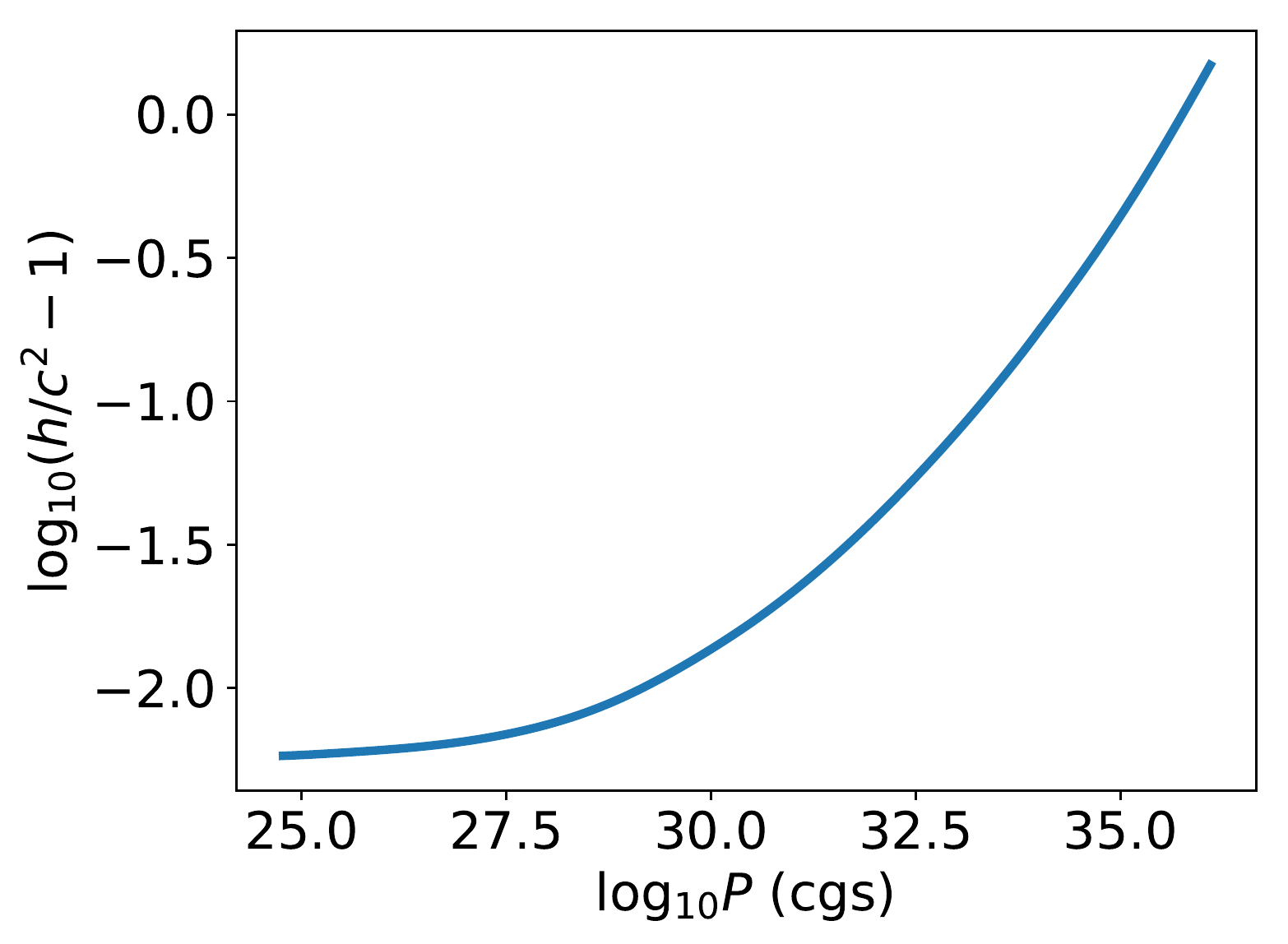}
  \caption{Specific enthalpy vs. pressure for $s=4$ $k_b/$baryon and
    $Y_e=0.1$. Note the offset along the y-axis. The specific enthalpy
    is not identically equal to the speed of light for small
    pressures.}
  \label{fig:s:contour:s15}
\end{figure}

\begin{table}[t!]
  \centering
  \begin{tabular}{l | r | r}
    \textbf{Parameter} & \textbf{Value} & \textbf{Units}\\
    \hline
    $M_{BH}$ & 2.80 & $M_{\odot}$\\
    $a_{BH}$ & 0.80 & $J_{BH}/M^2_{BH}$\\
    $s$ & 4.00 & $k_b/$baryon\\
    $Y_e$ & 0.10 & $n_e/(n_e + n_p)$\\
    $M_{\text{disk}}$ & 0.05 & $M_{\odot}$\\
    $R_{\text{in}}$ & 3.70 & $M_{BH}$ \\ 
    $R_{\text{max}}$ & 9.03 & $M_{BH}$ \\ 
  \end{tabular}
  \caption{Parameters of the initial data for our black-hole-disk
    system.}
  \tablecomments{$M_{BH}$ and $a_{BH}$ are the mass and spin of the black
    hole respectively. $s$ and $Y_e$ are the entropy and electron
    fraction, which are assumed to be constant throughout the initial
    disk. $M_{\text{disk}}$ is the mass of the disk. $R_{\text{in}}$ is
    the inner radius of the disk and $R_{\text{max}}$ is the radius of
    maximum pressure. We set $G=c=1$.}
  \label{tab:disk:params}
\end{table}

\begin{table}[t!]
  \centering
  \begin{tabular}{l | r | r}
    \textbf{Parameter} & \textbf{Value}\\
    \hline
    $N_1$ & 192\\
    $N_2$ & 128\\
    $N_3$ & 66\\
    $N_\nu/$cell & $5$\\
    $N_t$  & $1.6\times 10^6$\\
    $R_{\text{out}}$ & $10^3 M_{BH}$
  \end{tabular}
  \caption{Numerical parameters used for the post-merger disk
    calculation.}
  \tablecomments{$N_1$, $N_2$, and $N_3$ are the numbers of cells in
    the $r$, $\theta$, and $\phi$ directions, respectively.
    $N_{\nu}/$cell is the number of superneutrinos per cell.  $N_{t}$
    is the number of tracer particles.  $R_{\text{out}}$ is the radius
    of the outer boundary of the simulation with $G=c=1$.}
  \label{tab:disk:num:params}
\end{table}

We set up an axisymmetric disk in hydrostatic equilibrium on a Kerr
background.\footnote{We perform the setup in Boyer-Lindquist
  coordinates but transition to the modified Kerr-Schild coordinates
  described in \cite{HARM} for the evolution. For a detailed
  discussion of these various coordinate systems, see
  \cite{poisson2004relativist} and references therein.} We demand our
disk have constant fixed specific entropy $s$, specific angular
momentum $l$, and electron fraction $Y_e$. Under these conditions, the
relativistic Euler equations can be written as an exterior
differential system, which can be integrated along characteristics for
the specific enthalpy $h$ as derived in \cite{FishboneMoncrief}.

Figure \ref{fig:s:contours} shows the specific enthalpy as a function
of $\lrho$ and $\lT$ for the Hempel SFHo \cite{HempelEOS} equation of
state with fixed electron fraction $Y_e=0.1$. Overlayed on top of the
heatmap are contours of constant entropy. To construct a constant
entropy disk, we find one of these contours and move along it. Each
contour represents a relationship between $\lrho$ and $\lT$.

The exterior differential system for the enthalpy introduces several
constants of integration. These are set by the innermost radius of the
disk, $\Rin$, the radius of maximum pressure
$\Rmax$, and the limit
\begin{equation}
  \label{eq:disk:setup:h:bc}
  h_0(s,Y_e) = \lim_{P\to 0}h(s,Y_e),
\end{equation}
for a given entropy $s$ and electron fraction $Y_e$. For ideal gasses,
$h_0 \geq c$. However, this is not the case for more general equations
of state. The only constraint is that $h \geq 0$, as required by the
weak energy condition. We plot $h-c^2$ vs. pressure for $s=4$
$k_b/$baryon and $Y_e=0.1$ in figure \ref{fig:s:contour:s15}. Note the
offset along the y-axis.

We initialize the disk with a uniform, weak, purely poloidal magnetic
field with a minimum ratio of gas to magnetic pressure
\begin{equation}
  \label{eq:def:plasma:beta}
  \beta = 2\frac{P}{B^2},
\end{equation}
which acts as the seed field for the magneto-rotational instability
\cite{VelikovMRI}. We summarize our parameter choices for our disk in
table \ref{tab:disk:params}. We summarize the numerical parameters
used in table \ref{tab:disk:num:params}.

\subsubsection{Results}
\label{sec:disk:results}

\begin{figure}[tpb]
  \centering
  \includegraphics[width=\columnwidth]{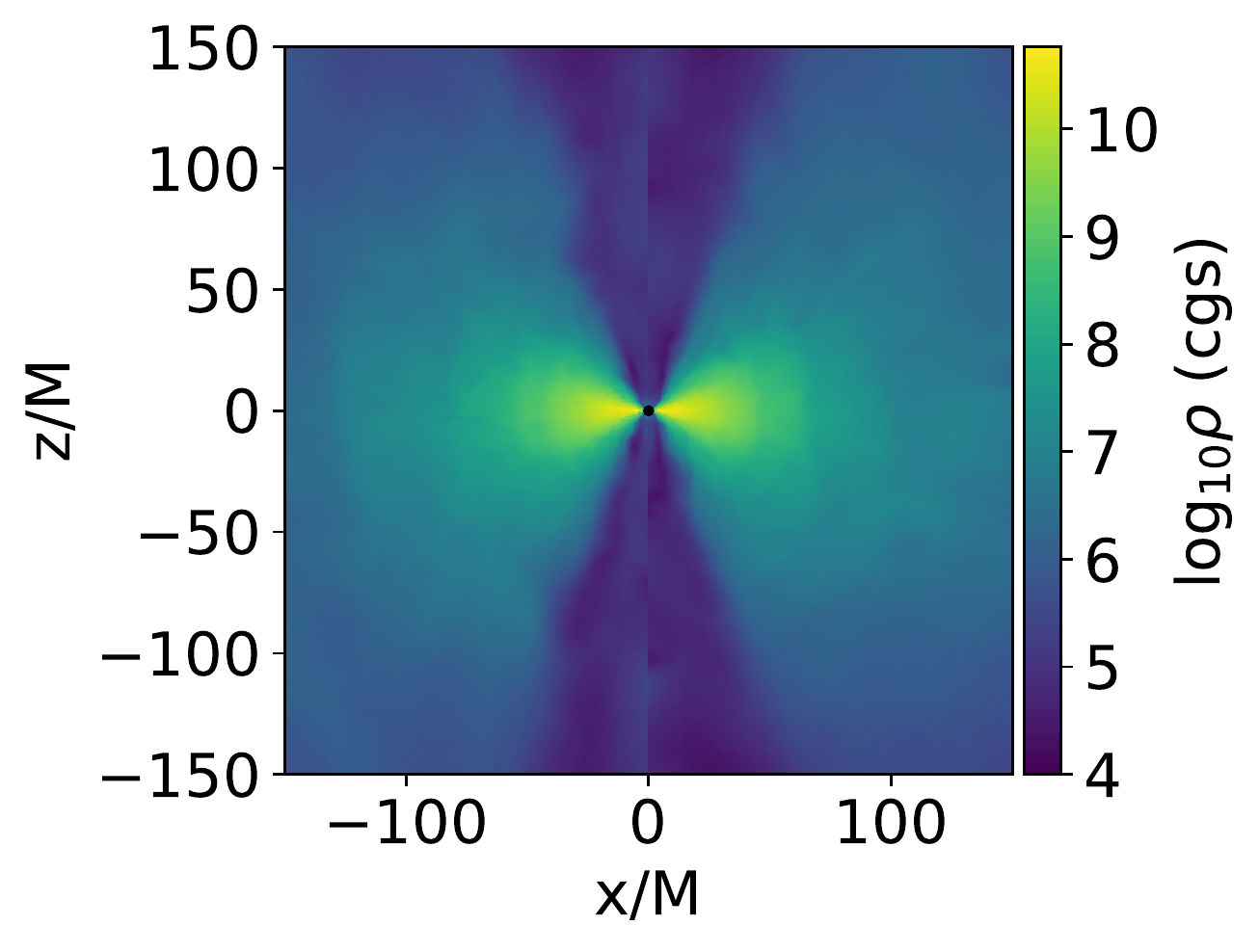}
  \caption{Density of the disk-outflow system in the $xz$-plane at
    time $10^4 \times G M_{BH}/c^3$.}
  \label{fig:disk:nu:wind:xz}
\end{figure}

\begin{figure}[tpb]
  \centering
  \includegraphics[width=\columnwidth]{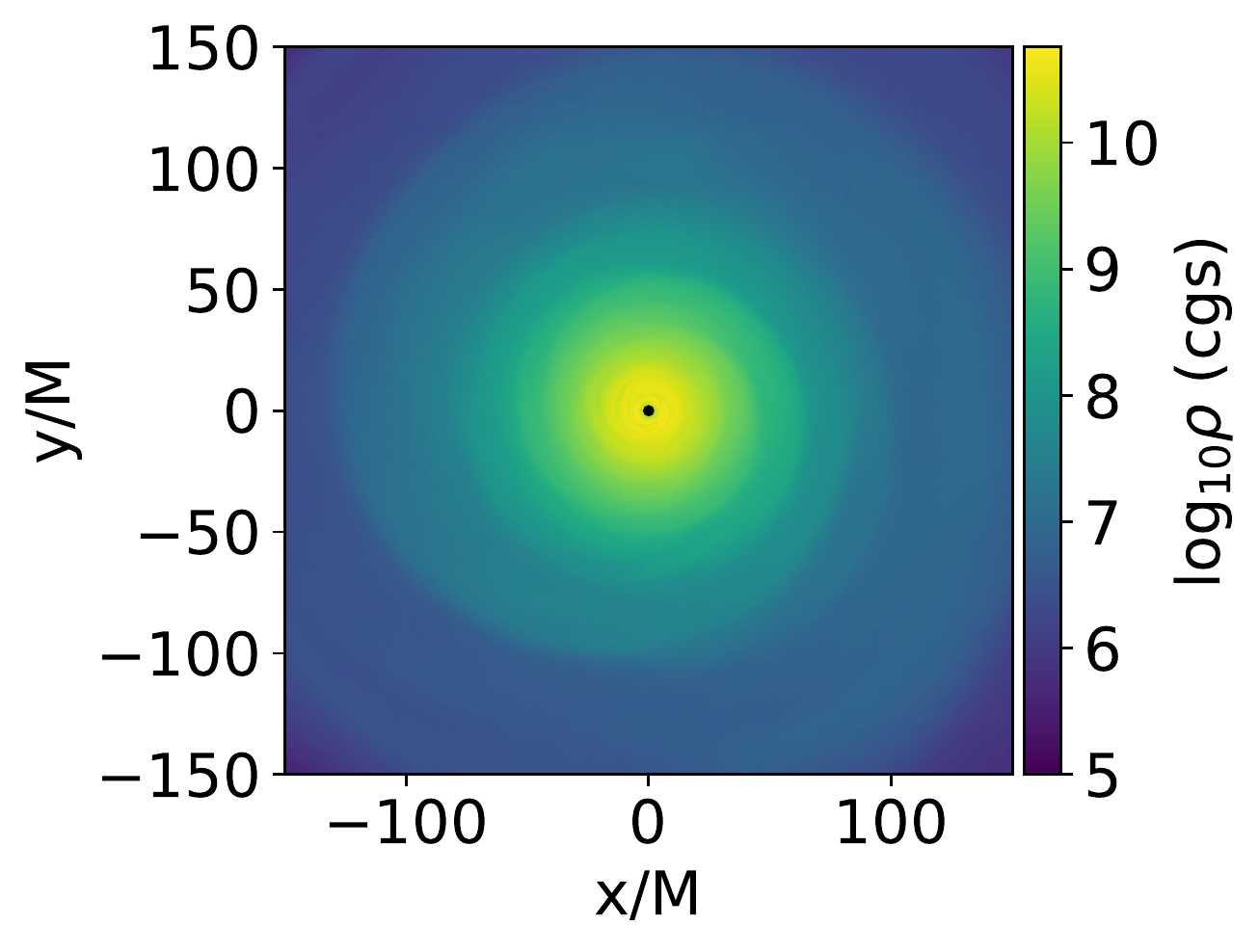}
  \caption{Density of the disk-outflow system in the $xy$-plane at
    time $10^4 \times G M_{BH}/c^3$.}
  \label{fig:disk:nu:wind:xy}
\end{figure}

\begin{figure}[tpb]
  \centering
  \includegraphics[width=\columnwidth]{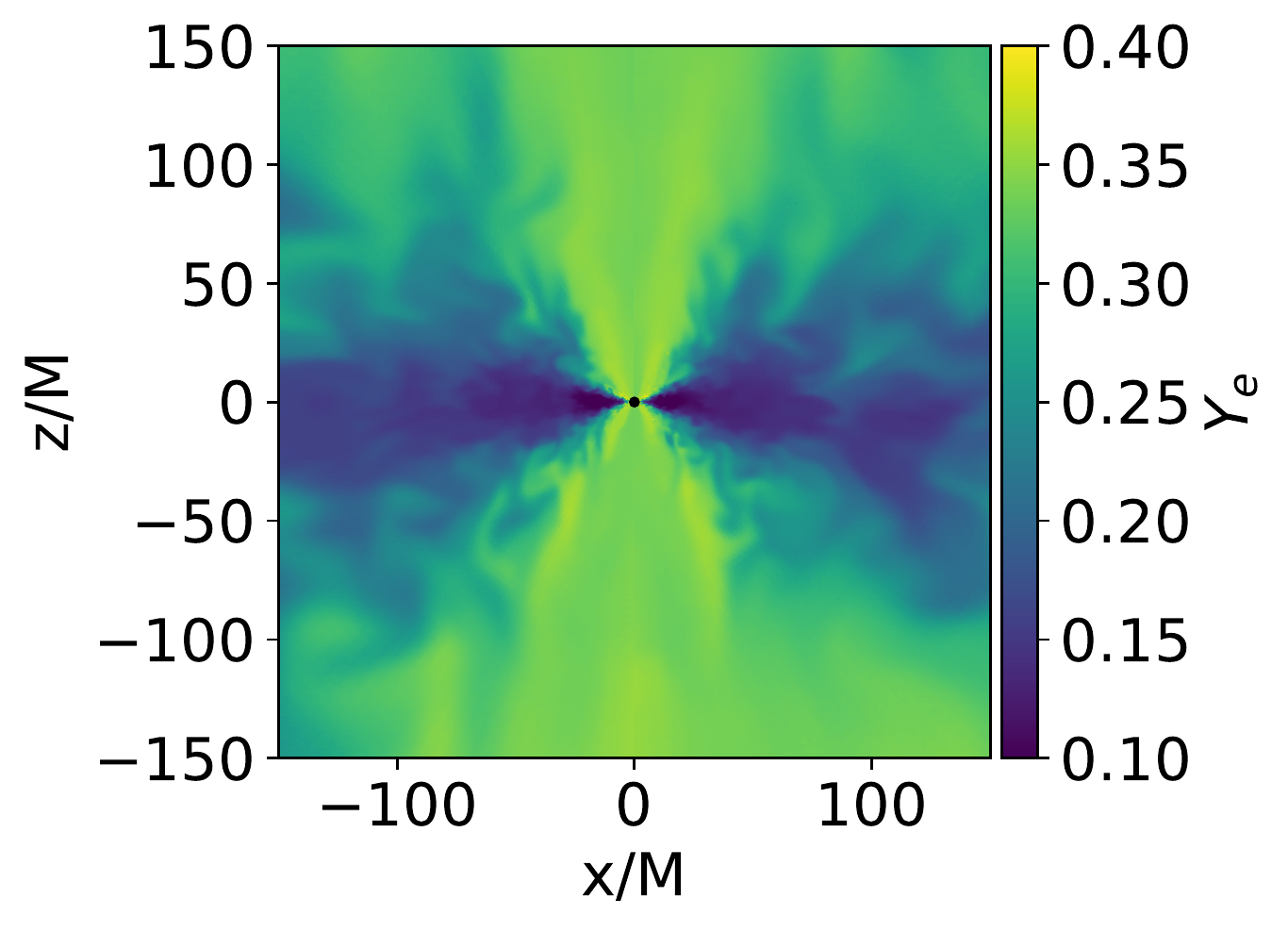}
  \caption{Electron fraction of material in the disk-outflow system in
    the $xz$-plane at time $10^4 \times G M_{BH}/c^3$. The composition
    of the material in the polar regions is untrustworthy, as this is
    jet material.}
  \label{fig:disk:wind:ye:xz}
\end{figure}

\begin{figure}[t!]
  \centering
  \includegraphics[width=\columnwidth]{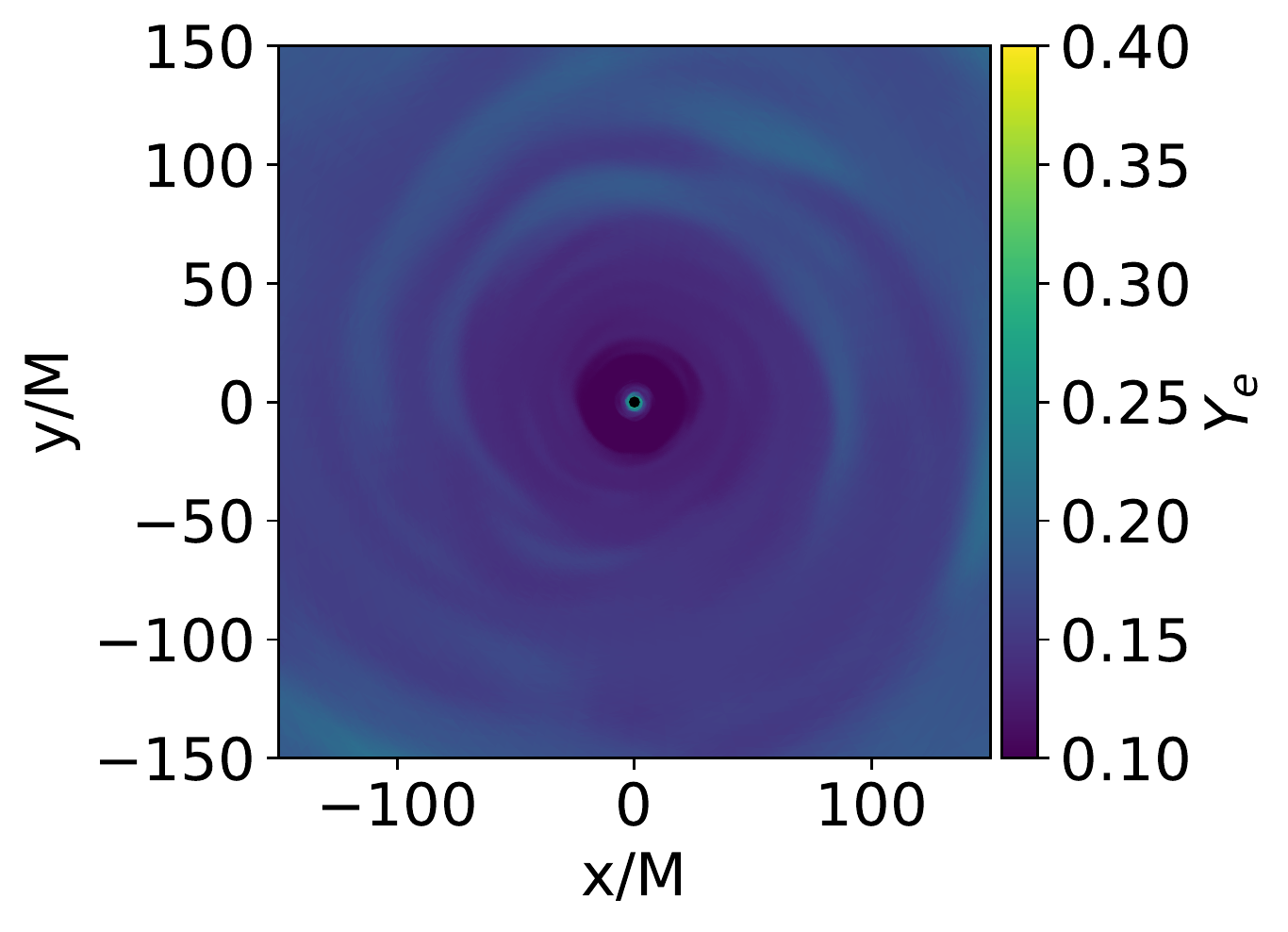}
  \caption{Electron fraction of material in the disk-outflow system in
    the $xy$-plane at time $10^4 \times G M_{BH}/c^3$.}
  \label{fig:disk:wind:ye:xy}
\end{figure}

We run our simulation for 10,000 $GM_{BH}/c^3$, or $\sim 138$
milliseconds. Figures \ref{fig:disk:nu:wind:xz} and
\ref{fig:disk:nu:wind:xy} show the density of the disk at late
time.. Neutrinos can carry lepton number and vary the electron
fraction as a function of space and time. Figures
\ref{fig:disk:wind:ye:xz} and \ref{fig:disk:wind:ye:xy} show the
electron fraction of the disk-wind system at late times. The core of
the disk still has very low electron fraction---close to
$Y_e\sim 0.1$. However, the outflow has a composition that varies
significantly in space. Material in the equatorial plane still has low
electron fraction, near $Y_e \sim 0.2$. However, material far from the
midplane has electron fraction as large as $Y_e \sim 0.3$.

\begin{figure*}[tpb]
  \centering
  \includegraphics[width=\textwidth]{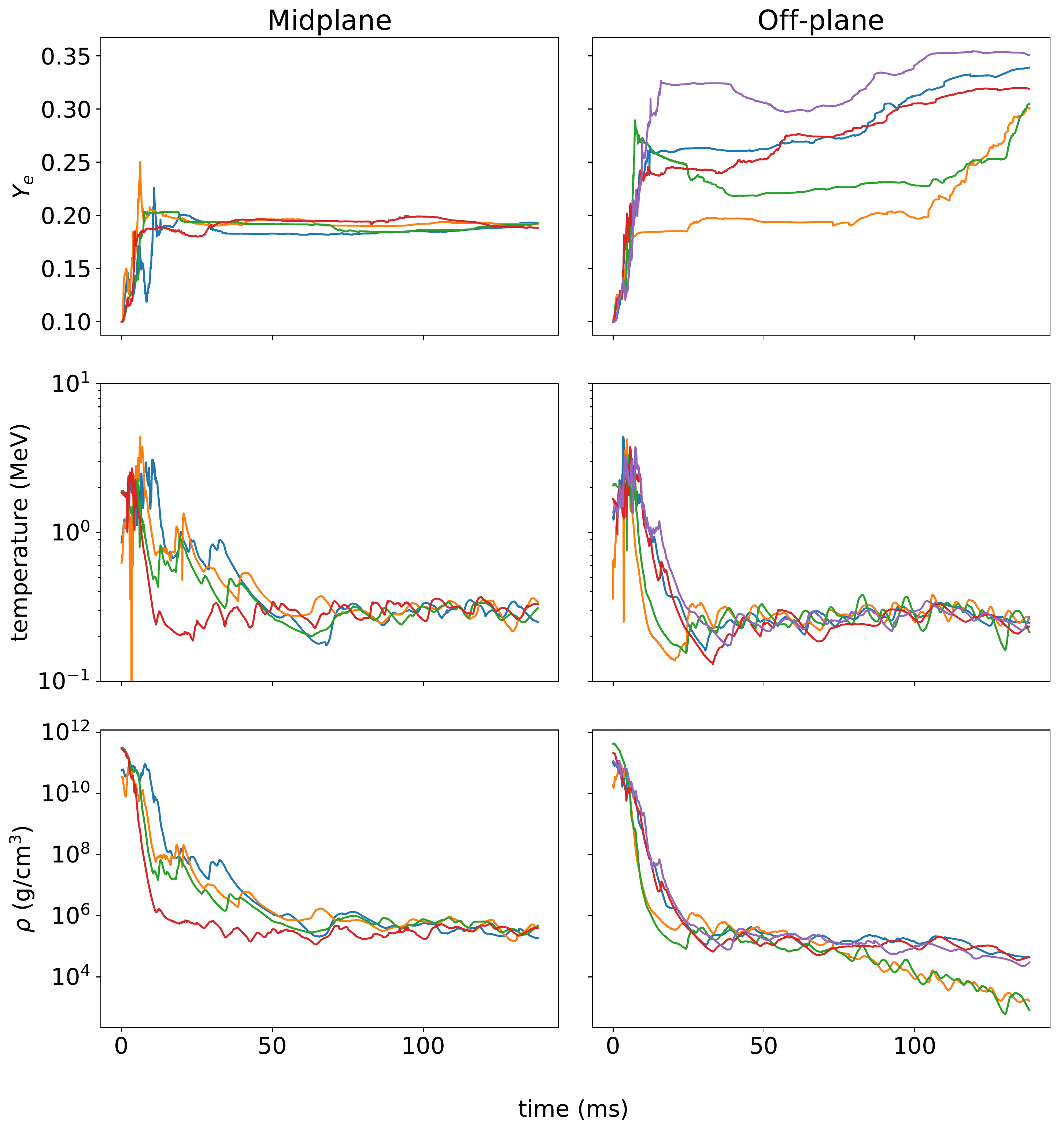}
  \caption{Density $\rho$, temperature $T$, and electron fraction
    $Y_e$ as a function of time for five characteristic tracers each
    for material in the midplane (left) and material near
    Boyer-Lindquist $\theta \sim \pi/3$ for tracers at $r \geq 100 M$
    at time $10^4 \times G M_{BH}/c^3$.}
  \label{fig:tracers:summary}
\end{figure*}

We make these qualitative observations more quantitative by examining
tracer data. For this analysis, we use tracers that have reached radii
greater than a fixed extraction radius $r_{min} = 100$ at time
$10,000 GM_{BH}/c^3$.
We plot density, temperature, and electron fraction as a function of
time for several characteristic tracer particles both in the midplane
and off-plane in figure \ref{fig:tracers:summary}. Although we do not
calculate yields here, these kinds of tracers may be used as input
into a nuclear reaction network to calculate yields. Although not
conclusive, we believe these results are highly suggestive that
realistic neutrino transport is required to accurately model these
systems. We will pursue this detailed modeling in future work.

\section{Concluding Thoughts}
\label{sec:conclusions}

We have developed the capacity to accurately study neutrino driven
accretion flows via Monte Carlo methods. We have validated the
accuracy of our approach via a variety of code tests, as discussed in
section \ref{sec:verification}. Moreover, we have demonstrated this
capability by performing a fully three-dimensional general
relativistic neutrino-radiation-magnetohydrodynamics calculation of a
representative black-hole-accretion-disk system formed by a compact
binary merger. With realistic neutrino transport active, we observe a
rich phenomenology of the wind morphology and composition.

Since we will observe many more compact binary mergers in the coming
years, accurately modeling these systems, and their remnants, is of
critical importance. We believe \nubhlight\ represents a key tool in
this modeling effort. In the future, we will use \nubhlight${}$ to
investigate the morphology of the disk-wind system in the context of
multimessenger observables.

\section{Acknowledgements}
\label{sec:ack}

The authors thank Chris Fryer, Francois Foucart, Daniel Siegel, Oleg
Korobkin, Jonas Lippuner, and Roseanne Cheng for many valuable
discussions. We are especially grateful to Adam Burrows for providing
us with neutrino opacity tables and explaining how these tables are
produced in \cite{BurrowsCorresp}.

We acknowledge support from the U.S. Department of Energy Office of
Science and the Office of Advanced Scientific Computing Research via
the Scientific Discovery through Advanced Computing (SciDAC4) program
and Grant DE-SC0018297

This work was supported by the US Department of Energy through the Los
Alamos National Laboratory. Additional funding was provided by the
Laboratory Directed Research and Development Program and the Center
for Nonlinear Studies at Los Alamos National Laboratory under project
number 20170508DR. Los Alamos National Laboratory is operated by Triad
National Security, LLC, for the National Nuclear Security
Administration of U.S. Department of Energy (Contract
No. 89233218CNA000001).

This research used resources provided by the Los Alamos National
Laboratory Institutional Computing Program, which is supported by the
U.S. Department of Energy National Nuclear Security Administration
under Contract No. 89233218CNA000001.

This article is cleared for unlimited release, LA-UR-19-20336.

We are grateful to the countless developers contributing to open
source projects on which we relied in this work, including Python
\cite{rossumPythonWhitePaper}, numpy and scipy \cite{numpy,scipyLib},
and Matplotlib \cite{hunterMatplotlib}.


\bibliography{nubhlight}

\end{document}